\documentclass[11pt]{article}   
\usepackage[letterpaper, left=.8in, top=.8in, right=.8in, bottom=.8in,nohead,includefoot]{geometry}
\usepackage{color,charter,graphicx,natbib,fancyhdr,bibentry,enumitem,here,setspace,multirow,amsmath,amssymb,amsfonts,subfig,placeins,mathdots,sidecap}
\usepackage{algorithm}\usepackage[noend]{algpseudocode}
\usepackage[svgnames,x11names]{xcolor}
\usepackage[pdftex]{hyperref}
\hypersetup{
colorlinks,%
linkcolor=MidnightBlue,  
urlcolor=DarkSlateGray,   
citecolor=RoyalBlue2  
}
\def\blu{\color{RoyalBlue4}}

\def\e{\textrm{e}}
\def\bx{\mathbf{x}}
\def\by{\mathbf{y}}

\def\bm{\mathbf{m}}
\def\ba{\mathbf{a}}

\def\bF{\mathbf{F}}

\def\bC{\mathbf{C}}

\def\bR{\mathbf{R}}

\def\bW{\mathbf{W}}

\def\bphi{\boldsymbol{\phi}}
\def\btheta{\boldsymbol{\theta}}

\def\bgamma{\boldsymbol{\gamma}}
\def\bomega{\boldsymbol{\omega}}

\def\cD{\mathcal{D}}
\def\cM{\mathcal{M}}

\DeclareMathOperator*{\argmax}{argmax}
\newcommand{\seq}[2]{#1\,{:}\,#2}
 
    \usepackage{titlesec} 
\titleformat*{\section}{\normalfont\Large\bfseries\blu }
\titleformat*{\subsection}{\normalfont\large\bfseries\blu }
\titleformat*{\subsubsection}{\normalfont\normalsize\bfseries\blu }
\titleformat*{\paragraph}{\normalfont\normalsize\bfseries\blu }
\titleformat*{\subparagraph}{\normalfont\normalsize\bfseries\blu }

\begin{document}

\setcounter{page}{0}\thispagestyle{empty}

\begin{center} 
{\LARGE\bf\blu Adaptive Variable Selection for Sequential Prediction  

\smallskip in Multivariate Dynamic Models}
 
\bigskip\bigskip
{\Large Isaac Lavine$^*$,   Michael Lindon$^\dagger$, and Mike West$^\diamond$}

\bigskip\bigskip
August 31, 2020

\bigskip\bigskip\bigskip\bigskip 

{\large\bf \blu Abstract}\end{center}

We discuss Bayesian model uncertainty analysis and forecasting in sequential dynamic modeling of multivariate time series. The perspective is that of a decision-maker with a specific forecasting objective that guides thinking about relevant models. Based on formal Bayesian decision-theoretic reasoning,  we develop a time-adaptive approach to exploring, weighting, combining and selecting models that differ in terms of predictive variables included.   The adaptivity allows for changes in the sets of favored models over time, and is guided by the specific forecasting goals.   A synthetic example illustrates how decision-guided variable selection differs from traditional Bayesian model uncertainty analysis and standard model averaging.  An applied study in one motivating application of long-term macroeconomic forecasting  highlights the utility of the new approach in terms of improving predictions as well as its ability to identify and interpret different sets of relevant models over time with respect to specific, defined forecasting goals. 
 
\medskip \noindent{\bf \blu Keywords}: Bayesian forecasting; Decision analysis;  Dynamic dependency network models; Dynamic linear models; Gibbs model probabilities; Macroeconomic forecasting; Model averaging; Model structure uncertainty; Shotgun stochastic search

\vfill
\footnotesize
\qquad 
\\
$^*$Susquehanna International Group LLP, PA, U.S.A.\href{mailto:isaac.lavine@duke.edu}{isaac.lavine@duke.edu}  
\\
$^\dagger$Netflix, CA, U.S.A.\href{mailto:michael.s.lindon@gmail.com}{michael.s.lindon@gmail.com}
\\
$^\diamond$Department of Statistical Science, Duke University, Durham NC 27708-0251, U.S.A., \href{mailto:mike.west@duke.edu}{mike.west@duke.edu}
\normalsize

\section{Introduction\label{introduction} }
 
Model structure uncertainty lies at the heart of much of scientific modeling but remains a central challenge to statistical methodology. Specific problems of variable selection and model weighting or averaging are central in Bayesian analysis and have seen enormous development to date.  However, more recent literature has increasingly emphasized the need for broader Bayesian views of model structure uncertainty. In particular, the proscribed nature of standard Bayesian model uncertainty and issues faced in realistic $\cM$-open settings have led to growing recent interest in new Bayesian approaches to scoring, weighting and combining models. We are concerned with these general questions in contexts of sequential analysis for forecasting and decision-making using dynamic state-space models for time series. Here the issues of model  uncertainty  are exacerbated by the potential for relevant, data-respecting models to change in structure over time, as well as for parameters within a model to be time-varying. This, together with broader challenges,  has been highlighted across a class of dynamic model contexts in \cite{West2020Akaike}.   The current paper picks-up the theme and addresses the challenges with a novel Bayesian approach to weighting and, over time, adaptively reweighting structures from large model classes, guided by specific forecasting and/or decision goals. 

Critical limitations of model probabilities are significantly highlighted in the sequential time series setting. First, model marginal likelihoods 
score only $1-$step ahead forecasting accuracy. The marginal likelihood value on a model from $n$ observations is the product of realized values of $1-$step forecast densities.  This clearly demarks the applied relevance of this score.  Models are built with forecasting and decision goals, and $1-$step forecast accuracy is rarely the only motivating goal.  A model scoring highly in that sense may be hopeless for multi-step ahead forecasting,  or define poor forecasts for resulting decisions.  More broadly, the need to  consider explicit goals in model structure assessment  has been recognized at least implicitly in recent literature on model weighting and combination~\citep[e.g.][]{ClydeIversen2013,Amisano2017,McAlinnWest2017bpsJOE,McAlinnEtAl2017,McAlinnEtAldiscussionBA2018,YaoGelman2018}, and explicitly in some areas related to multi-step forecasting and decisions when comparing and combining models~\citep[e.g.][]{NakajimaWest2013JBES,Kapetanios2015,West2020Akaike}.   A model that forecasts well on one subset of outcomes in a multivariate setting may be poor in other dimensions. A model scoring highly on one purely statistical metric  may be inferior to other models in a decision problem or in terms of contextually relevant forecast accuracy measures~\citep[e.g.][]{BerryWest2018DCMM,BerryWest2018TSM}.  We argue for a more explicit, core focus on integrating forecasting and decision goals as arbiters of model assessment to advance  practically relevant methodology. 

A further concern is that practical interests in model uncertainty rarely include identifying \lq\lq true models''; rather, model structure is often  a   nuisance parameter and not of inherent interest otherwise. In variable selection,  identifying a model, or a few models, that are useful for prediction is typically the goal.  The academic enterprise of treating increasingly large sets of models defined by many subsets of potential predictors quickly runs into the well-known-- and intractable-- problems of model multiplicities, redundancies and collinearities: many models with differing structures generate similar predictions, collinearities drive complications in interpretation, and model averaging induces increased noise in resulting predictions~\citep[e.g.][]{Hans2007,EdGeorge2010,GiannoneEtAl2018}. 
Practically,  interest often lies in \lq\lq  good choices'' in terms of forecast and decision outcomes~\citep[e.g.][]{GruberWest2016BA,GruberWest2017ECOSTA,West2020Akaike}.   Then, the increasing dimension of model spaces
argues against the traditional Bayesian view of maintaining interest in all possible models.  In sequential analysis of time series this is particularly highlighted, as monitoring and updating scores on many models over time quickly raises the computational stakes. As many models will be of little or no interest,  coupled with the common issue of huge redundancy of model classes,  this argues for selective analysis of smaller numbers of models and a concern to-- at selected points over time-- review and refresh selected sets of models under consideration.  

We address the above issues with a new Bayesian approach to adaptive (over time) model uncertainty analysis. The ideas are general while being presented in the motivating context of multivariate time series forecasting with specific forecast goals, and in which the model structure in question is the specification of sets of predictor variables in dynamic linear models for the multivariate series.   The example context uses flexible classes of dynamic dependency network models for a vector time series, and explores analysis in a topical macro-economic forecasting context.  Section~\ref{timeseries} defines the time series setting. 
Section~\ref{avs} opens with  explicit desiderata underlying the perspective on sequential analysis and forecasting in the context of predictor variable uncertainty,
responding to the issues and challenges discussed above. This section then defines both the conceptual basis and technical/computational details of the novel 
adaptive variable selection strategy.  A simulation study in Section~\ref{simulationstudy} is followed by results from a macroeconomic case study in Section~\ref{macroeconstudy}.  The application focuses  on the relevance of model structure uncertainty with respect to multi-step ahead and path forecasting, i.e., the specific and key goal in monetary policy-related contexts of forecasting trajectories of economic indicators over a path of time points into the future. Summary comments appear in Section~\ref{conclusions}, with supporting material in Appendices.

\section{Time Series Setting and Perspectives \label{timeseries}}

\subsection{Multivariate Time Series: Notation and Models} 

The $m\times 1$ vector $\by_t$ comprises a set of $m$ univariate time series $y_{j,t}$ in equally-spaced time.
The class of Dynamic Dependency Network Models (DDNMs) is a flexible framework for modeling and forecasting, and is increasingly  exploited due to the ability to customize univariate series, sensitively model cross-series relationships and their dynamics over time, and to scale with $m$. DDNMs couple together sets of univariate dynamic linear models (DLMs) and exploit the well-known, analytic forward filtering and forecasting results of DLMs~\citep[e.g., chapt. 4 in each of][]{WestHarrison1997,PradoWest2010}. Full details can be seen in~\cite{ZhaoXieWest2016ASMBI}, with a recent, relevant example in~\cite{IrieWest2018portfoliosBA}.  The cross-series structure of DDNMs is also intimately related to other popular multivariate models applied in economics, finance and related areas~\citep[e.g.]
[]{Primiceri05,NakajimaWest2013JBES,NakajimaWest2013JFE,ZhouNakajimaWest2014IJF,NakajimaWest2015DSP,NakajimaWest2017BJPS,Shinichiroetal2017,Lopes2018}. 

A DDNM is defined by a set of univariate dynamic models 
\begin{equation} \label{eqn-univariate-dlm}
y_{j,t} =\bF'_{j,t} \btheta_{j,t} + \nu_{j,t}, \quad \bF_{j,t} = \begin{pmatrix} \bx_{j,t} \\ \by_{pa(j),t}\end{pmatrix}, \quad
 \ \btheta_{j,t}=  \begin{pmatrix}\bphi_{j,t} \\ \bgamma_{j,t}  \end{pmatrix}, \quad \nu_{j,t} \sim N(0,1/\lambda_{j,t}),
\end{equation}
where $j=\seq{1}m$ indexes series and  $t=1,2,\ldots$ indexes time. In each series $j,$ the state vector and volatility $(\btheta_{j,t},\lambda_{j,t})$ evolve via a  linear state equation coupled with a discount volatility model,  assumedly independently across series.
Observation errors $\nu_{j,t}$ are independent across $j$ and over $t$. 
The regression vector $\bF_{j,t}$  involves: (a)  a subvector $\bx_{j,t}$ of exogenous predictors and/or selected lagged values of some of the $m$ series-- giving opportunity for sparse and time-varying vector autoregressive components as well as external predictor variables; (b)  a subvector $\by_{pa(j),t}$ of  {\em parental predictors}-- here $pa(j) \subseteq 
 \{ \seq {j+1}m \}$ is an index set selecting some 
  of the contemporaneous values of other series ordered higher than $j$ in the vector.  
 The triangular structure of the parental sets defines the multivariate model of $\textbf{y}_t$ by a series of conditional relationships.  The conformably partitioned state vector  $\btheta_{j,t}$ includes subvectors of dynamic coefficients  $\bphi_{j,t}$ on exogenous and lagged predictors, and $ \bgamma_{j,t} $ on parental predictors, viz
 $E(y_{j,t}|\ast) = \bF'_{j,t} \btheta_{j,t}  = \bx'_{j,t} \bphi_{j,t} + \by'_{pa(j),t} \bgamma_{j,t} $ where $\ast$ indicates all relevant terms.
 
The joint distribution can theoretically be decomposed into any series ordering; in practice the series order matters for interpretation and will impact on variable selection for each of the decoupled univariate models.    In our economic time series examples below we follow prior authors in choosing a contextually relevant ordering~\citep[e.g.][]{NakajimaWest2013JFE,Eickmeier2015}. For broader commentary on the ordering in applications, we refer to the particularly germane discussion and reply to discussion in~\cite{ZhaoXieWest2016ASMBI}, as well developments in related models in~\cite{NakajimaWest2015DSP} and~\cite{Crespo2019}. Common empirical experiences are that the
impact of the ordering is typically very limited from the viewpoint of forecasting accuracy. 
Otherwise, with interests in  justification a modeller is free to define her/his own ordering based on application-specific rationales, and then explore and evaluate multiple possible choices.    While these are important applied considerations, they are not of main interest in connection with the primary contributions of this paper and we proceed with a chosen order.

At each time $t,$ denote by $\cD_t$ the current information set. This includes initial information $\cD_0,$ all data $\by_1,\ldots,\by_t$ up to time $t$, and all other information 
used in the modeling process-- including the $\bx_{j,t}$,   interventions or changes to model structure, etc.  Implicitly in what follows, $\cD_t$  also includes information needed or relevant in forecasting multiple steps ahead, including future values of exogenous predictors.

\subsection{Sequential Learning and Forecasting \label{learningDLMs}} 

The models of eqn.~(\ref{eqn-univariate-dlm}) are standard DLMs amenable to analytic computation for forward filtering and $1-$step ahead forecasting. 
In our example context,  the evolution of $\btheta_{j,t}$ is a simple (linear, conditionally normal) random walk, and  is coupled with  a discount/random walk volatility model for $\lambda_{j,t}$. This standard framework allows for change over time controlled by discount factors~\citep[e.g.][section 4.3]{PradoWest2010}. A lower discount factor allows more substantial changes over time, while a discount factor of 1 corresponds to static coefficients; a brief summary of time $t$ evolution and updating appears in Appendix 1 
of the Supplement.  Importantly, filtering analyses are both analytic and conditionally independent across series $j$ so are done in parallel, while forecasting involves recoupling across series.  

At time $t$ for each series $j$, the conditional (on parental predictors) $1-$step ahead forecast distribution $p(y_{j,t} | \by_{pa(j),t}, \cD_{t-1} ) $ is a univariate T-distribution with trivially computed parameters.  This yields the joint forecast density function via composition, i.e.,  
$p(\by_t | \cD_{t-1}) = \prod_{j=1}^m p(y_{j,t} | \by_{pa(j),t}, \cD_{t-1}).$
For $k>1,$ forecasting $k-$steps ahead is   based on direct simulation, exploiting the recursive structure of DDNMs. This enables computationally trivial simulation of the path of the multivariate time series over the next $k$ time points.    Technically, this simply propagates samples of the paths of states and volatilities $(\btheta_{j,\ast}, \lambda_{j,\ast})$  for each series $j$, coupled with sampling from the conditionally normal DLMs to generate the $\by_\ast.$  Standing at time $t$, for example, this evaluates the full path forecast distribution  by generating Monte Carlo samples, a.k.a. \lq\lq synthetic futures'', from 
$p(\by_{t+1},\ldots,\by_{t+k} | \cD_t) = \prod_{h=1}^k p(\by_{t+h} | \by_{t+1},\ldots,\by_{t+h-1}, \cD_t).$   All practical forecasting interests over the coming $k$ periods can then be addressed with relevant Monte Carlo summaries (e.g, expected or median paths,  prediction of turning points, maxima or minima,  value-at-risk, expected utility functions, etc.)


\subsection{Model Uncertainty:  Predictive Variable Specification   \label{modeluncertaintyDLMs}} 


The central model structure uncertainty question in DDNMs is specification of predictor variables in both exogenous/lagged $\bx_{j,t}$ terms and parental sets $pa(j).$  
Write   $\cM_j$ for a set of candidate models $\cM_j^r$ for series $j,$  indexed by $r\in \{ \seq 1{|\cM_j|}\}$.   
  Mathematically, $\cM_j^r \in \{0, 1\}^p$ is a $p-$dimensional vector selecting predictor variables.
An important point is that we will be expanding the framework so that model spaces are effectively time dependent, i.e., $\cM_j \to \cM_{j,t}$, but for now maintain the simpler notation. 
Denote by $\cM$ any single multivariate model for $\by_t$ defined by a selection of one model from each of the $m$ sets $\cM_j.$ 

Consider first the discount learning extension of standard Bayesian model probability analysis, and BMA as a special case.
At any time $t-1$, denote the current model probabilities by $p(\cM | \cD_{t-1}).$ Given a model space discount factor $\alpha$ such that $0<\alpha\le 1,$ the discount modified model probability at time $t$  is defined by 
$$
p(\cM  | \cD_t) \propto  p(\cM| \cD_{t-1})^\alpha  \ p(\by_t | \cM, \cD_{t-1}) 
 \propto p(\cM|\cD_0 )^{\alpha^t}\  \prod_{h=1}^t p(\by_h | \cM,\cD_{h-1})^{\alpha^{t-h}}
$$
where the earlier notation for $1-$step forecast p.d.f.s has been extended to be explicit that it depends on the specific model structure $\cM.$   A discount $\alpha<1$ acts to reduce the impact of historical information in model comparisons,  with data from $n$ time points in the past discounted by $\alpha^{n}$ in the cumulation of model scores.   Evidently, standard Bayesian analysis sets $\alpha=1.$ 
As $t$ increases, $\alpha<1$ means that model weights will not degenerate;  they adapt over time and respond to varying $1-$step predictive abilities across the sets of models~(\citealp[][p.445]{WestHarrison1989}; \citealp{Raftery10}; \citealp{Xie2012}; \citealp{Koop2013}; \citealp{ZhaoXieWest2016ASMBI}). A major potential benefit is that of adapting more rapidly to reweight models based on more \lq\lq local'' behavior in the series, and down-weight models that were historically more favored but are locally of lower predictive value.  This often yields improved predictive performance as illustrated in multiple examples in   the above references.

A major benefit of DDNMs is that model uncertainty is addressed across series $j$ independently, as dynamic variable selection problems in each of the univariate DLMs.  This implies  $ \sum_{j=1}^m |\cM_j|$ possible models $\cM$, whereas a direct multivariate analysis would involve a much more substantial set of $\prod_{j=1}^m |\cM_j|$ models.    That is, as earlier noted, 
$
p(\by_t | \cM,\cD_{t-1}) 
  \propto  \prod_{j=1}^m   p(y_{j,t} | \cM_j,\by_{pa(j),t} \cD_{t-1})$
so  the contributions to model scores given by the set of $m$ $1-$step forecast p.d.f. values are decoupled across series.  
 
Traditional Bayesian analysis-- perhaps with the practically motivated but otherwise subjective intervention-based discount model probability variant-- proceeds using model scores defined above. 

\section{Time-Adaptive Variable Selection} \label{avs}

\subsection{Model Structure Uncertainty and Practical Forecasting \label{desiderata} } 

Following discussion and motivation in Section~\ref{introduction}, we develop analysis consistent with the following perspectives. 
\begin{itemize} \itemsep-3pt
\item Performance in prediction with respect to specific, defined forecasting goals should arbitrate model comparisons, combination and selection. 
Evaluation of alternative models, and the definition of models scores to use in weighting models for aggregation in prediction and selection of future \lq\lq optimal'' models,  should consider specific forecast and/or decision goals.  
\item At each time $t,$  it is desirable to have a single chosen model for communication and use in forecasting,  and changes to the chosen model over time justified based on improvements in forecast accuracy modulo specific forecasting goals.  
\item Consideration of banks of models to assess any \lq\lq current'' model, and combination of selected sets of models for forecasting purposes, should be entertained at any times that forecast accuracy under that chosen model might be questioned. This can be done routinely   at each time point, or at selective time points based on model monitoring and assessment of predictive accuracy modulo the specific forecasting goals~\citep{West1986,West1986a,West1989,GruberWest2016BA,GruberWest2017ECOSTA}.
\end{itemize}  
The methodological contributions of this paper include a  strategy for time-adaptive variable selection that address these desiderata.  The resulting 
adaptive variable selection (AVS) strategy is composed of:  (1)  so-called Gibbs model probabilities, tying model evaluations with defined forecasting  objectives;  (2) a local search strategy over model spaces to dynamically explore potential models relative to a \lq\lq current'' selection;  (3) a choice of a representative model at each time point for communication,  
interpretability and as a basis to evolve forward in time; and (4) the use of (1-3) adaptively over time.

\subsection{Gibbs Model Probabilities \label{GibbsModelProbs} }

Our approach relates to the growing interest in Bayesian decision-guided inference with loss or utility functions used to define mechanisms to update subjective probabilities over models (or, more generally, over uncertain states and parameters).  We use the term \lq\lq Gibbs model probabilities'', contributing to the growing literature concerned with so-called generalized belief updating in which data-based evidence is represented in likelihood functions constructed based on defined loss or utility functions~\citep{Tanner2008,BissiriWalker2016}.  Previous work has used purely statistical loss functions, and established that such an approach can provide superior risk performance to Bayesian updating under model misspecification~\citep{Zhang2006a, Zhang2006b, Tanner2008}. 
Beyond expanding the ideas to dynamic model structure uncertainty  and developing a sequential, adaptive approach, a key focus here is to exploit the approach using loss or utility functions specific to the main prediction problems of interest, also linked to prior work on explicitly recognizing model selection as a decision~\citep[e.g.][]{HahnCarvalho2015}.

Consider series $j$ with  (time $0$) baseline model probabilities $p(\cM_j^r |\cD_0 )$ over selected models $\cM_j^r \in \cM_j$.  Gibbs model probabilities based on   data $\cD_t$ (from all $m$ series) observed up to time $t$ are defined by 
\begin{equation} \label{eq:Gibbsmodelprobsseriesj}
p_j(\cM_j^r | \cD_t) \propto p(\cM_j^r |\cD_0 ) \e^{\tau  s_{j,t}(\cM_j^r)}  
\end{equation}
where $\tau>0$ and $s_{j,t}(\cM_j^r)$ is a {\em model score}.  A higher model score indicates more support for the   model $\cM_j^r$ and reflects  historical performance   in a specific forecasting or decision problem. The scores are defined by choosing a utility function relevant to the specific goals. Examples include simple point forecast metrics or full (log) predictive densities for functions of the outcome time series as used in our examples below.  With scores on a known or standardized scale, the
parameter $\tau$ balances information from the past data with that in the prior. Questions of how to calibrate $\tau$ are discussed in~\cite{BissiriWalker2016} and in 
our settings in Sections~\ref{simulationstudy} and~\ref{macroeconstudy}  where scores are based on out-of-sample predictive densities.    The general setting allows scores to be of different forms across series $j,$  or to be be based on a common function but with series $j-$specific weights, or to involve the same score for each component series.  Examples below use the latter, while the generality is important to note for future applications.     Gibbs model probabilities are used for model averaging just as in standard model uncertainty analysis; note that the latter arises as a special case when $\tau=1$ and scores are simply the logs of $1-$step ahead predictive densities. More generally, the Gibbs likelihoods can be interpreted as extensions of model likelihood from the $1-$step ahead predictive focus  to that based on a   desired forecast or decision goal.

One of the major benefits of DDNMs is that, as discussed in Section~\ref{modeluncertaintyDLMs}, marginal likelihoods for a multivariate model are simply the products of likelihoods from each of the $m$ univariate models.  This carries over to Gibbs model probabilities assuming that the scores are unrelated  and that baseline models are independent across series.  Then the overall probabilities on the multivariate model $\cM$ defined by the set of $m$ chosen models $\cM_j^{r_j}$ is simply the product of terms in eqn.~(\ref{eq:Gibbsmodelprobsseriesj}), 
\begin{equation} \label{eq:Gibbsmodelprobsseriesmultivariate}
p(\cM | \cD_t )   \propto p(\cM | \cD_0 )   \e^{\tau  s_t(\cM)},  \qquad s_t(\cM) =   \sum_{j=1}^m  s_{j,t}(\cM_j^{r_j})
\end{equation}
where $ p(\cM | \cD_0 )  $ is the product of the $p(\cM_j^r |\cD_0 ). $ 
The overall probabilities $p(\cM | \cD_t )$ are used for model averaging for prediction and decisions, and then model selection. 
That is,  model evaluation is decoupled to the levels of the univariate series,   then recoupled to assess the overall multivariate model.

\subsection{AVS Strategy and Representative Model Selection \label{AVSrepresentative} } 

The overall strategy of adaptive variable selection (AVS) is summarized in Algorithm 1 below. At each time $t,$ we find a set of models using Shotgun Stochastic Search (SSS), a strategy to explore regions of strong models \citep{Jones2005a,Hans2007,ScottCarvalho2008,HaoWang2015}. Models are evaluated using Gibbs model probabilities, and averaged together for forecasting and decisions before moving on to the next time point. The AVS strategy is run in parallel over series $j=\seq{1}m$ to find candidate models and evaluate their Gibbs probabilities. Forecasting from multivariate DDNMs is then trivial via sequential simulation.

\begin{algorithm}  \label{alg-avs}
	\caption{Adaptive Variable Selection}
	\begin{algorithmic}[1]
		\For {time t in 1:T}
		\For {series j in 1:m}
		\State Find a set of candidate models $\cM_j$ with SSS, seeded by representative model $\cM_{0, t-1}$
		\State Calculate Gibbs probabilities $p_j(\cM_j^r | \cD_{t-1})$
		\EndFor
		\State Forecast with model averaging, where the probability of DDNM $\cM$ is the product of its univariate model probabilities.
		\State Observe $y_{t}$, and for each series $j$, update Gibbs Probabilities $p_j(\cM_j^r | \cD_t)$
		\State Choose a new representative model $\cM_{0, t}$
		\EndFor
	\end{algorithmic}
\end{algorithm}

  As discussed in Section~\ref{introduction} and the desiderata of Section~\ref{desiderata},  it is often desirable for interpretation and communication to operate using a single selected model unless or until changes are suggested based on a breakdown in model performance or external considerations.   Thus selecting one model as a representative of the probability-weighted set is of interest.    Denote by $\cM_{0,t}$ a DDNM chosen as the {\em representative model} at time $t.$  A natural choice is the modal model with respect to Gibbs model probabilities, i.e., modulo the baseline probabilities that model maximizing the overall score $s_t(\cM).$    Alternatives would choose $\cM_{0,t}$ as a Bayesian decision with respect to the mixture over models.  A natural  approach would choose the representative model to best approximate (e.g., using Kullback-Leibler divergence) a specific predictive distribution that averages over the full set of models under consideration. This has  theoretical and practical appeal, but is computationally expensive relative to selection of the modal model.

Analysis proceeds through the DDNM evolution to time $t+1$ using the single model $\cM_{0,t}.$  
Then observing $\by_{t+1}$ we face the question of identifying classes of models $\cM_j$ and computing Gibbs model probabilities. The theoretical indication that we continuously update scores on all possible models is simply not practicable in realistic settings.  Then,  as time evolves different models become of interest relative to those that had scored well in the past. Further, interventions at certain times may change the class of models under consideration (e.g. by adding new potential predictors not so far considered). Hence the interest is (a) to identify sets of models at time $t+1$ that appear competitive with $\cM_{0,t}$ in terms of the specific forecasting goals, i.e., in terms of the defined score function, 
while (b) to do so computationally efficiently as this will be repeated at each time point.   Our AVS implementation utilizes an extension  
of shotgun stochastic search (SSS) to address these goals, as detailed in Section~\ref{SSS} below. 

A practical modification is to use the model search and weighting via AVS only at selected time points.  That is,  at time $t+1$ and over a number of further time points,   we may simply use the single model $\cM_{0,t}$ for evolution and forecasting.  At some point, however,  consideration of other models will become important, and then the AVS strategy of finding and weighing sets of models will come into play.

\subsection{Finding Models: Shotgun Stochastic Search \label{SSS} }

Originally developed for graphical models and regression, shotgun stochastic search  is designed to quickly identify and explore interesting regions in large, discrete model spaces~\citep{Jones2005a,Hans2007,ScottCarvalho2008,HaoWang2015}. \nocite{Hans2007b}
Its proven ability to rapidly transit model spaces based on \lq\lq local changes'' to existing models makes it perfectly suited to the AVS context in DDNMs with larger numbers of potential predictor variables per series.  At time $t+1$, the current representative model $\cM_{0,t}$  serves as an initial \lq\lq seed model''.  Based on this, SSS proceeds as follows:

\begin{enumerate}

	\item Identify a neighborhood of the seed model, typically the set of models \break $\{\cM_{0, t}^{+}, \cM_{0, t}^{\circ}, \cM_{0, t}^{-}\}$, where
		\begin{enumerate}
			\item $\cM_{0, t}^{+}$ is all models with 1 predictor added,
			\item $\cM_{0, t}^{\circ}$ is all models with 1 predictor swapped,
			\item $ \cM_{0, t}^{-}$ is all models with 1 predictor subtracted.
		\end{enumerate}
	In DDNMs this applies separately to each of the $m$ decoupled DLMs for univariate series.

\item Evaluate all such models in the neighborhood.  This can use posterior model probabilities or Gibbs model probabilities, or any other scoring method desired (e.g., scores from specific decision problems--~\citealp[e.g.][Section 2]{West2020Akaike}).
\item Record this set of models and scores in a running list.
\item Sample a new seed model from this neighborhood, and repeat. Sampling will be done using model probabilities or the decision-guided Gibbs extensions. 
\end{enumerate}
When the seed model is highly scoring, then the set of neighboring models will typically include many other interesting models in terms of the score. SSS therefore fully exploits local modes in model space to swiftly move between individual high probability models to reach varied parts of the model space.  
Neighboring models can be evaluated in parallel, which is a clear advantage over sequential search methods. This makes SSS particularly suited for situations where full exploration of the model space is computationally impossible, either because the set of models is large, or calculating scores is slow.   Importantly, the goal is to identify subsets of highly scoring models to underlie forecasting and evolution to the next time point; the goal is explicitly quite different to that of MCMC-based model search strategies, i.e. of \lq\lq structure learning''.    

Within the SSS search at each time, each model identified requires fitting over a period of past data-- possibly all data from $t=0$ or perhaps over a restricted recent period-- to evaluate model scores based on the historical forecasting record.  That DDNMs admit fast, analytic computation is critical here, enabling evaluation of even very large sets of candidate models at each time point; again, these computations are inherently decoupled hence parallelizable within each time point.

\section{Synthetic Time Series Example} \label{simulationstudy}

\begin{figure}[hbpt!]
\centering
\includegraphics[width=11 cm]{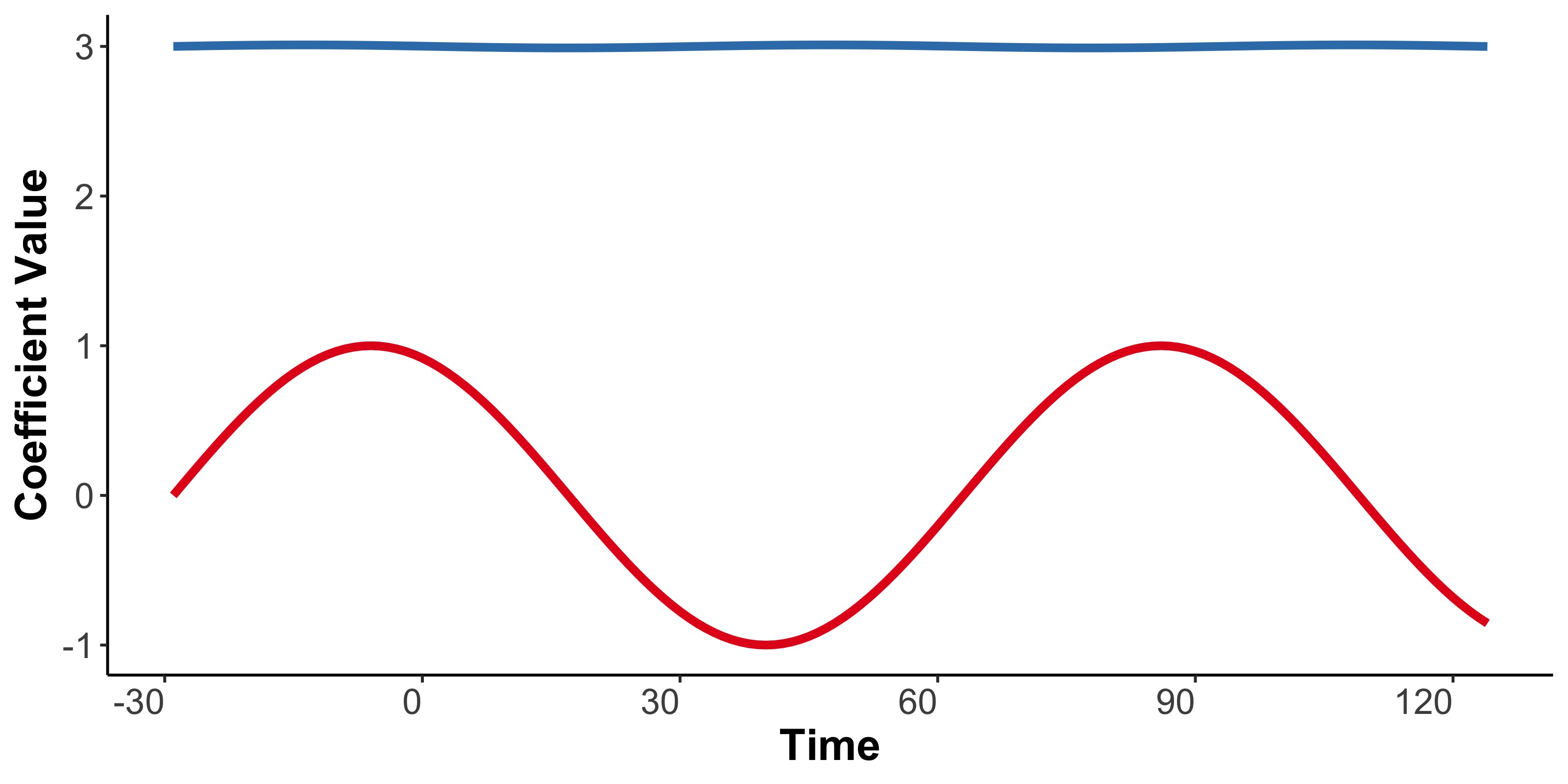}
\caption{Time-varying coefficients $\theta_{1,t}$ ({\em red}) and $\theta_{2,t}$ ({\em blue}) underlying synthetic data. 
\label{fig-thetas}  }
\end{figure}

A simple but relevant and illuminating example with synthetic data illustrates AVS compared to standard model averaging, and demonstrates how AVS selects predictors with stable effects for long-term forecasting. This is clearest in the simple case where $m=1$ series so the DDNM reduces to a single DLM at $j=m=1,$   with data $y_t \equiv y_{1,t}.$   
The data are simulated from a model exhibiting both steady and more rapidly changing dynamics. We generate $y_t = c + \theta_{1,t} x_{1,t} + \theta_{2,t} x_{2,t}$, with simulated $\theta_{1,t}$ and $\theta_{2,t}$ displayed in Figure~\ref{fig-thetas}. Note that $\theta_{1,t}$ is rapidly changing, while $\theta_{2,t}$ is relatively steady.  Predictors $x_{1,t}$ and $x_{2,t}$ are randomly set at $1$ or $-1$ with probability $1/2$.

Each model $\cM$ is a univariate DLM defined by a choice of predictor variables. All models include an intercept so there are 4 possible combinations of the variables $x_{1,t}$ and $x_{2,t}$ for inclusion, defining 4 candidate models at each time.  In each DLM, the state vector and volatility processes follow standard random walk evolutions as earlier discussed, with discount factors $\delta = \beta = 0.98$; see also Appendix 1 
of the Supplement. 
Gibbs model probabilities use $\tau=1$ as scores are log forecast densities so the resulting probabilities are on the same scale as standard Bayesian model probabilities.  The baseline priors at $t=0$ give equal weight to each model, and conjugate normal/inverse gamma priors for the state vector and volatility in each model are based on informal analysis of 
data from an additional training period of 30 time steps before formal model scoring and AVS analysis begins over $t=100$ time points.

\begin{figure}[htbp!]
\centering
\includegraphics[width=10 cm]{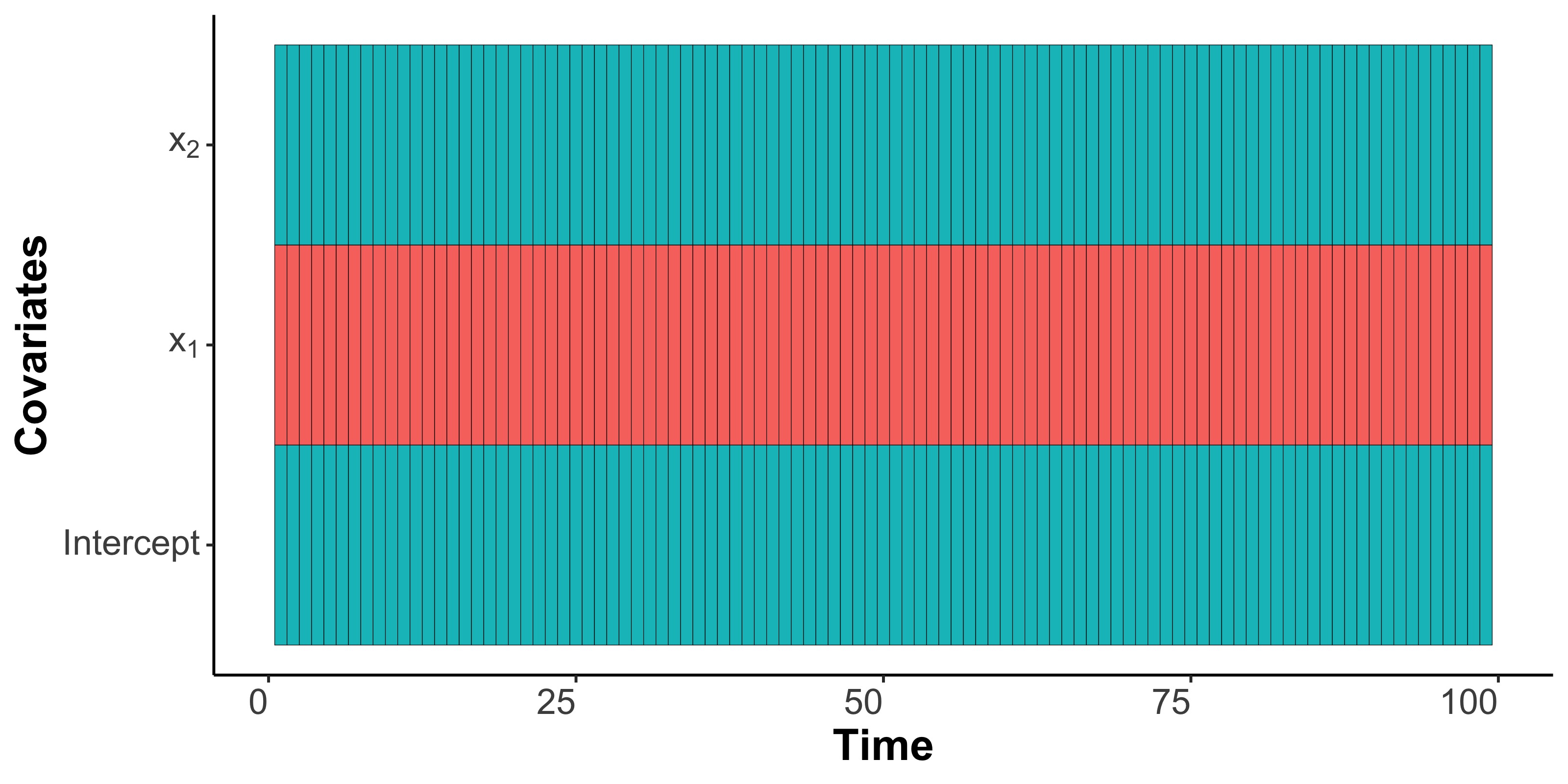}
\includegraphics[width=10 cm]{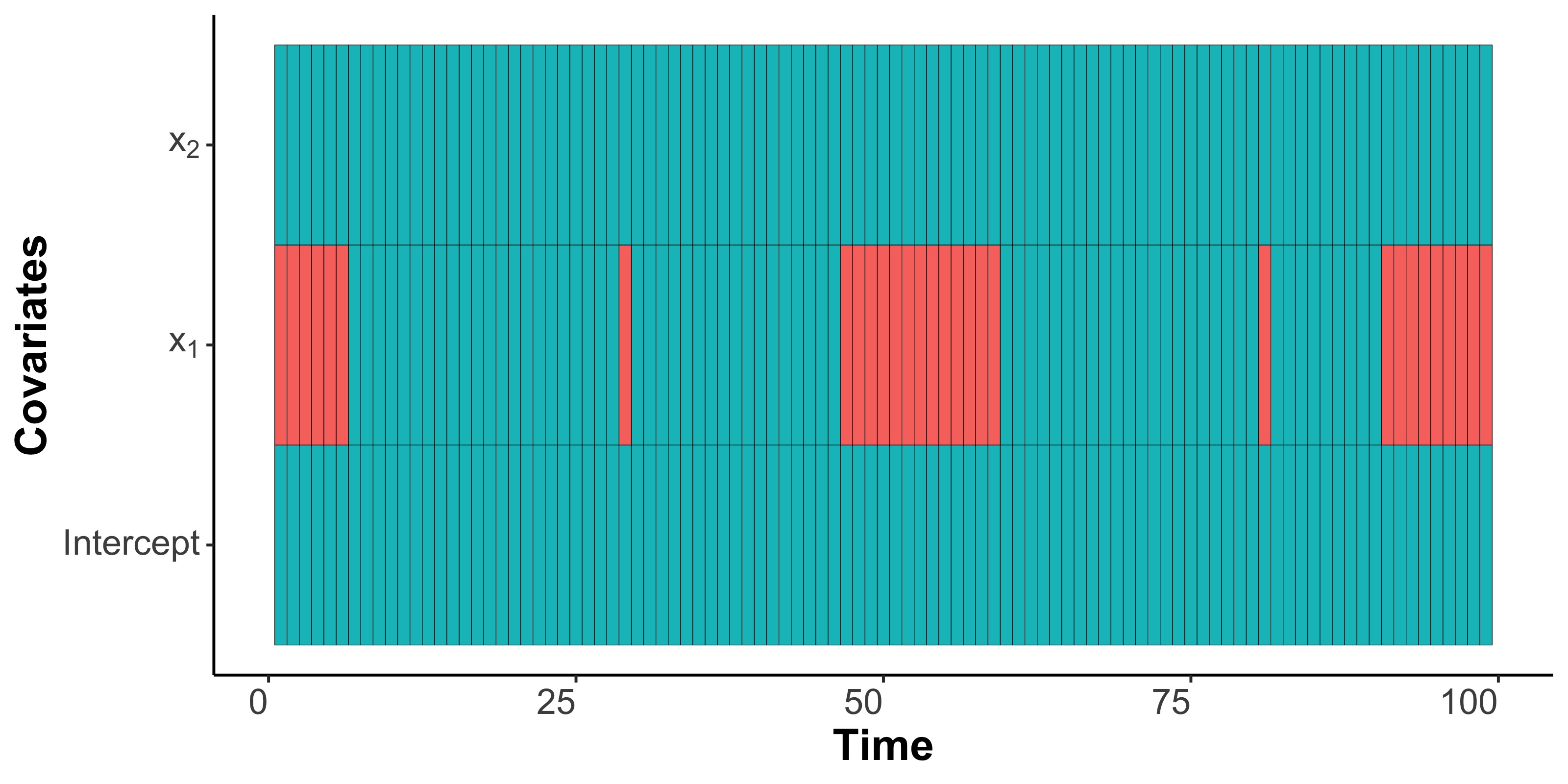}
\caption{Synthetic data example: Indicators of inclusion ({\em green}) of $x_{1,t}$ and $x_{2,t}$ in the posterior modal models under AVS ({\em upper}) and 
BMA ({\em lower}).  \label{fig-sim_avs_bma_model} }
\end{figure} 

\newpage
 
\begin{figure}[htpb!]
\centering
\includegraphics[width=10 cm]{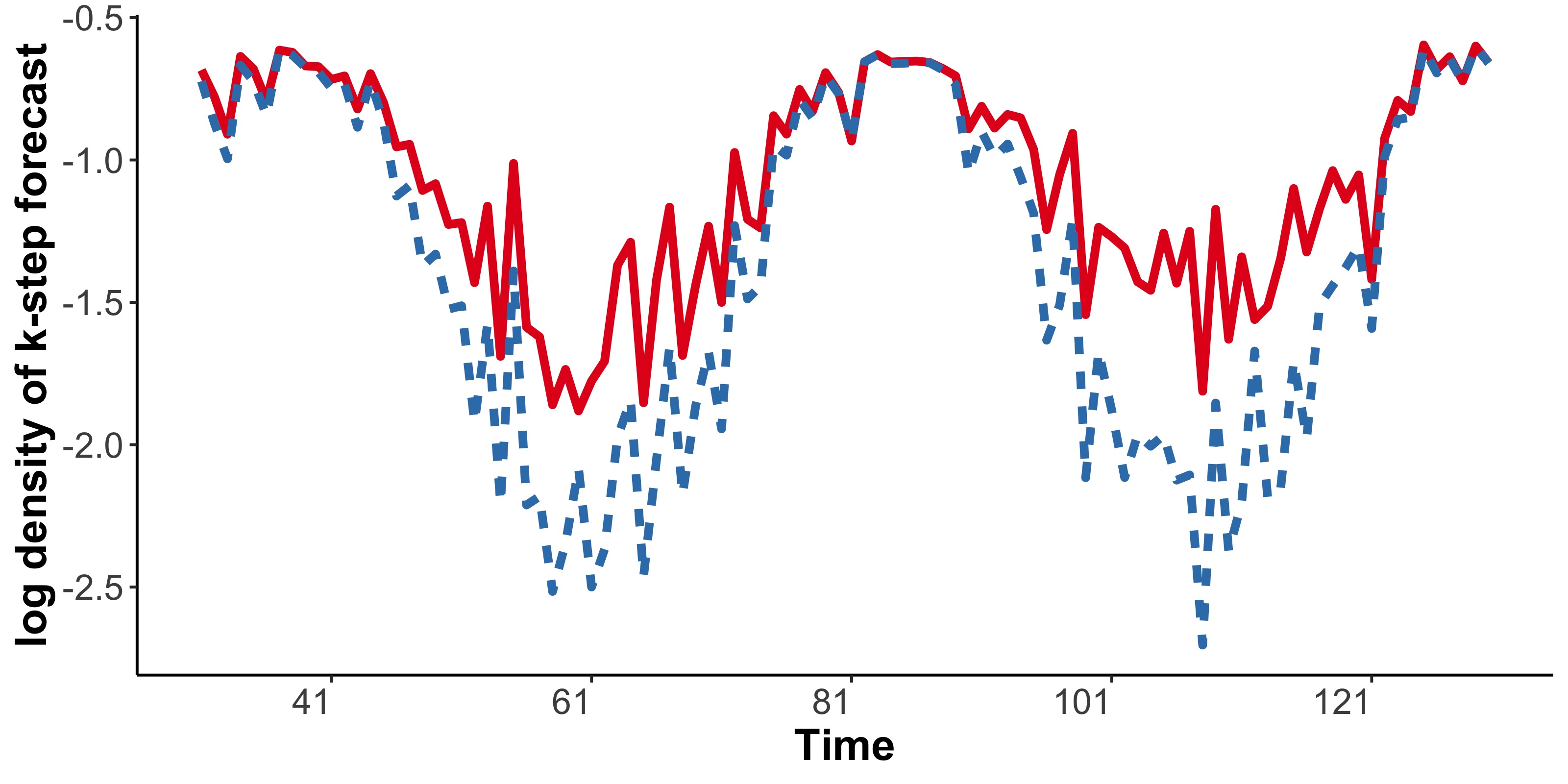}
\includegraphics[width=10 cm]{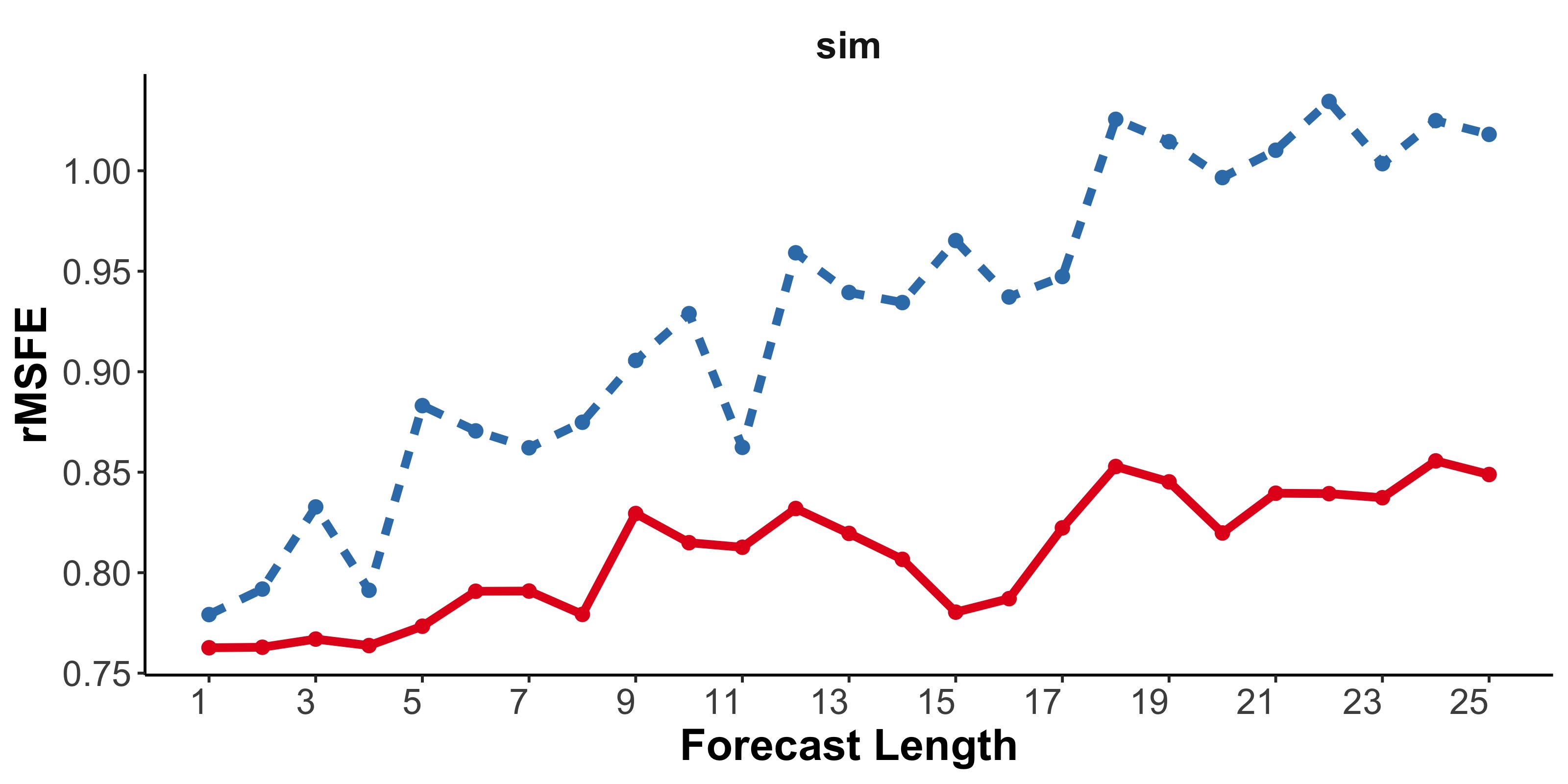}
\caption{Synthetic data example: Multi-step ahead forecast performance using AVS ({\em red, solid}) (with $k=25$) and BMA ({\em blue, dashed}). Log forecast density scores 
$\log (p(y_{t+25}|\cD_t))$ over time $t$  ({\em upper}), and root mean squared forecast errors as a function of forecast horizon ({\em lower}).
\label{fig-sim_evals} }
\end{figure}

Figure~\ref{fig-sim_evals} shows that AVS forecasting dominates BMA in terms of model-averaged predictive density.   With $k=25$ to drive AVS,    the $k-$step ahead based  predictive density score naturally improves over the myopic BMA.    Smaller differences occur at periods when the BMA drops $x_{1,t}$ or, by chance, $\theta_{1, t} \approx \theta_{1, t+k}$.   More deeply, using the same AVS analysis with $k=25$ in fact improves forecast accuracy over all horizons, as exhibited by the marginal root mean squared forecast error (rMSFE) for each horizon $1-25$ steps ahead in the figure.  This occurs even though the model is weighted by $k=25-$step ahead scores only, and the figure highlights the fact that standard BMA will tend to perform well only at short horizons. 

Reflecting central interests in multi-step ahead forecasting arising in many applications~\citep[e.g.][]{NakajimaWest2013JBES}, Gibbs model probabilities at each time are based on model scores of marginal $k-$step ahead forecast accuracy with $k=25$ as an example.  The model score function  $s_t(\cM) \equiv s_{1,t}(\cM_1)$ on each model $\cM$ is simply
$$
s_t(\cM) = \sum_{h=0}^{t-k} \alpha^{t-k-h} \log (p(y_{h+k} |\cM, \cD_h) )
$$
for some model discount factor $\alpha\in (0,1]. $  Here $\alpha<1$  down-weights more distant past outcomes as in the discount Bayesian model uncertainty analysis of Section~\ref{modeluncertaintyDLMs} that arises as the special case when $k=1;$  standard BMA is given with $\alpha=k=1.$  Our example here sets $\alpha = 0.98$ for both Gibbs and Bayesian model probabilities. Previous studies have identified $0.95 < \alpha < 1$ as an appropriate range, with little benefit in considering multiple $\alpha$ values ~\citep{Raftery10, Koop2013, ZhaoXieWest2016ASMBI}.

The behavior of adaptive variable selection is best illustrated through the identification of the representative model, taken here as the 
Gibbs posterior modal model at each time point; see Figure~\ref{fig-sim_avs_bma_model}.  Predictor $x_{1,t}$ is uniformly excluded; inclusion of a variable with rapidly changing and unpredictable dynamics generally degrades long-term predictions. In contrast, the posterior modal model from BMA almost always includes $x_{1,t}$, except when the coefficient $\theta_{1,t}$ drops near to $0$.

\section{Macroeconomic Case Study} \label{macroeconstudy}

 \subsection{Forecasting Context and Data} 
\begin{figure}[htbp!]
\centering
\includegraphics[width=10 cm]{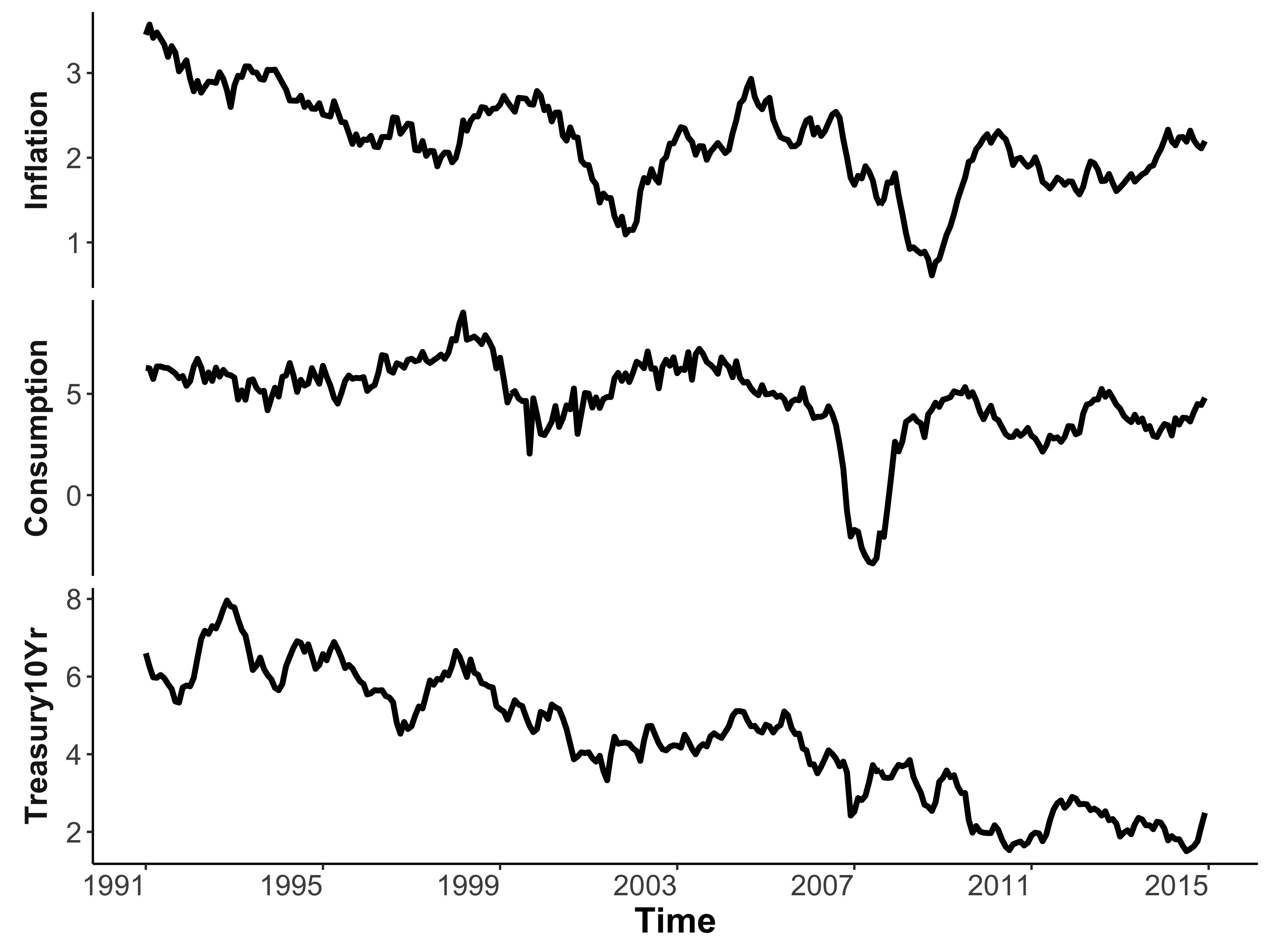}
\caption{Macroeconomic example: Monthly time series of US Inflation, Consumption, and yield on 10-year Treasury bonds (Tr10Yr) over 24 years up to the end of 2016.
\label{fig-data} }
\end{figure}

We address monthly forecasting of three key US macroeconomic series: year-over-year Inflation, Consumption, and the 10-year yield on Treasury bonds (Tr10Yr). 
Data over $1991-2016$ from the St. Louis Federal Reserve are shown in Figure~\ref{fig-data}. The sharp drop in both Inflation and Consumption during recessions is clear in 2001 and 2008, while Inflation and Tr10Yr show slight long-term downward trends.    Improved forecasting of these and related series is a central concern in national monetary policy, and forecasting more than a few months ahead is notoriously challenging (e.g. ~\citealp{Primiceri05}).   While 
particular interests lie in forecasting $12-24$ months ahead at each time point, central bank concerns spread across forecast horizons. It is becoming increasingly clear that customizing models to the forecast horizon of interest can improve forecast accuracy and potentially generate economic insights into dynamic relationships among series over time~\citep{NakajimaWest2013JBES,McAlinnWest2017bpsJOE,McAlinnEtAl2017}.  

\bigskip

Potential predictors include all  $1-12$ month lags of each  series. The DDNM orders series as  Inflation-Consumption-Tr10Yr.  Hence  Consumption and Tr10Yr are potential parents of Inflation,  Tr10Yr is a potential parent for Consumption, while Tr10Yr has no parents. Including a possible intercept, the total predictor space  has 39 potential predictors  for Inflation, 38 for Consumption, and 37 for Tr10Yr.  We summarize  forecasting results from analyses as follows. Earlier data from $1986 - 1990$ was used informally to choose informative priors at the start of 1991 for all states and volatilities. Analyses were run for a training period of 5 years, and then full forecast evaluations were made over the 252 month period $1996-2016$ inclusive. 

\subsection{Horizon-specific Multi-step Forecasting \label{kstepmargin} }

\begin{figure}[h!]
\centering
\includegraphics[width=10 cm]{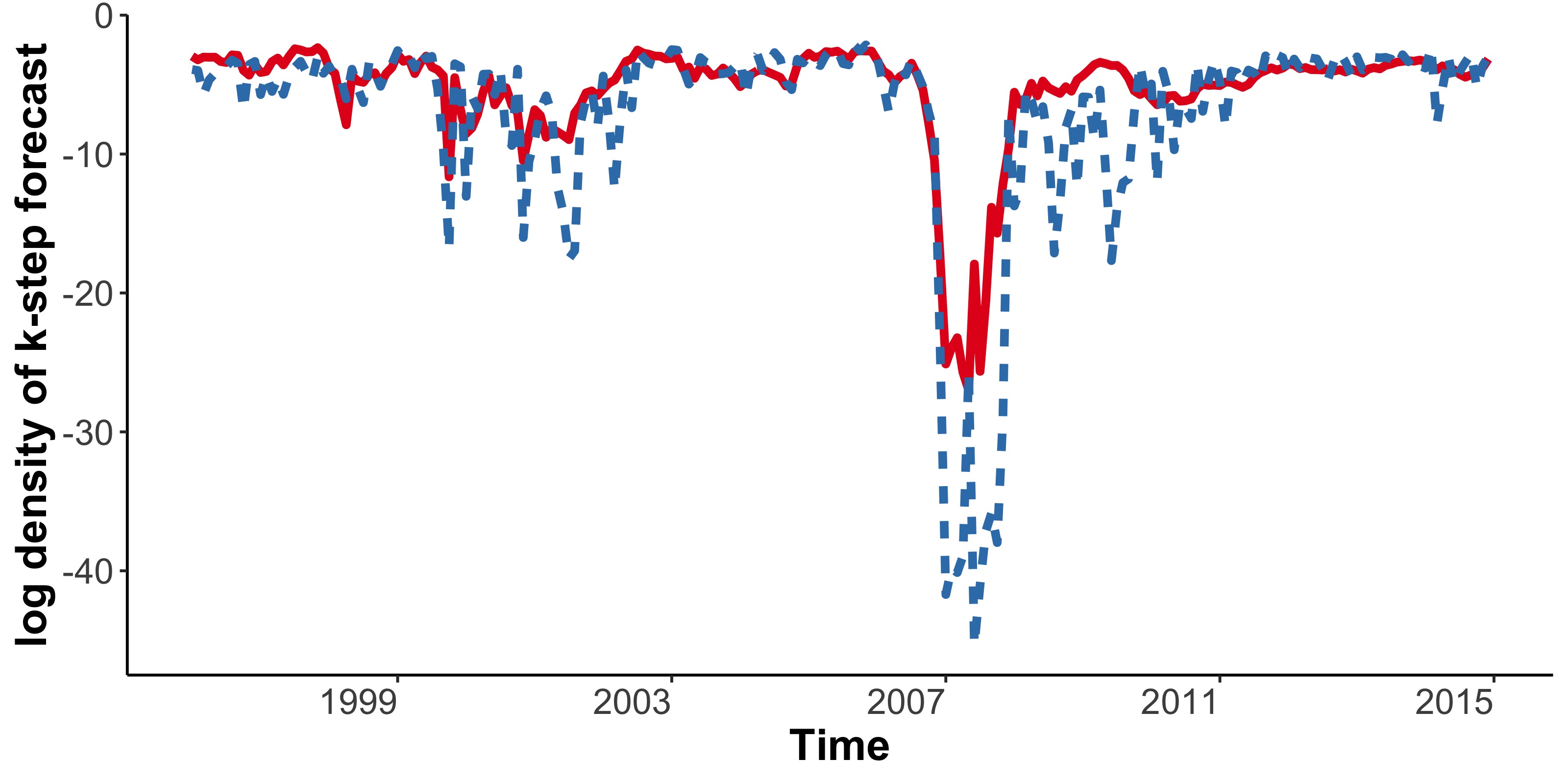}
%
\includegraphics[width=10 cm]{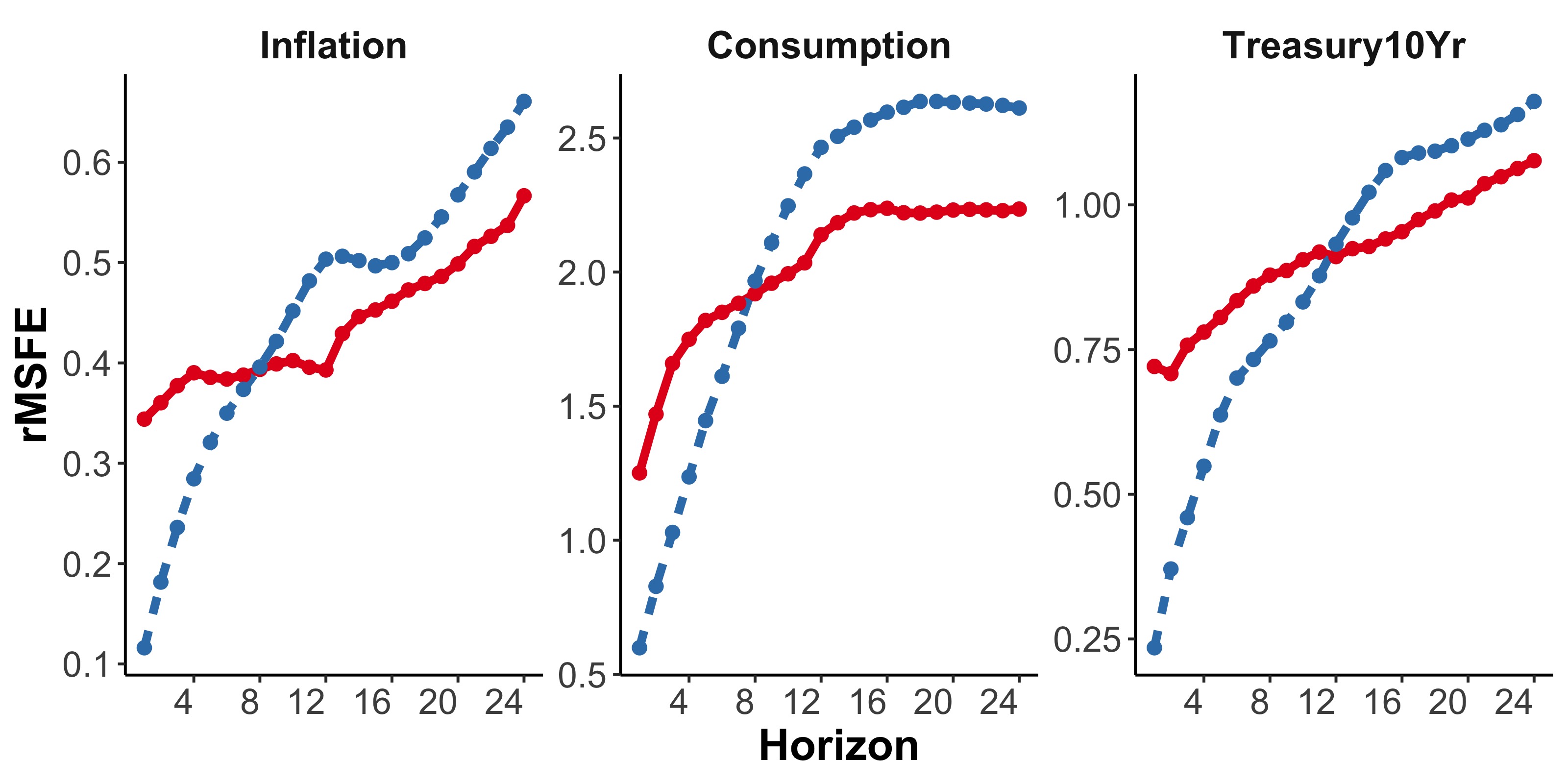}
\caption{Macroeconomic example:  Marginal $k=24-$steps ahead forecasting comparisons between AVS ({\em red, solid}) and BMA ({\em blue, dashed}).  Log forecast densities $\log (p(\by_{t+24}|\cD_t ))$ over time $t$ ({\em upper}), and marginal root mean squared forecast errors over 1 to 24 month forecast horizons ({\em lower}).
\label{fig-t_plus_k_forecastperformance} }
\end{figure}

Initial analysis considers marginal forecasts for $k=24$ months ahead.  The score is the discounted  log predictive density at the chosen horizon as in the univariate example in Section~\ref{simulationstudy} but now for the multivariate series; that is,  
\begin{equation}\label{eq:kstepforecastdensity} s_t(\cM) = \sum_{h=0}^{t-k} \alpha^{t-k-h}  \log p(\by_{h+k} | \cM, \cD_h) )
\end{equation}
where $\by_{h+k}$ is the multivariate observation at time $h+k$, and both AVS and BMA use $\alpha = 0.98$ for this study. Note that the computational cost of AVS scales {\em linearly} in the number of series because we are able to take advantage of the DDNM structure. Model scores are   evaluated for each series $j$ independently, and in parallel, by conditioning on the past values of other series. Forecasts are evaluated using the joint $24-$step ahead forecast density. Figure~\ref{fig-t_plus_k_forecastperformance} shows that AVS dominates BMA with respect to the long-term forecasting objective function defined by the usual model-averaged predictive density while, as expected, BMA analysis is more accurate in the shorter-term predictions.

\begin{figure}[h!]
\centering
\includegraphics[width=.75\textwidth]{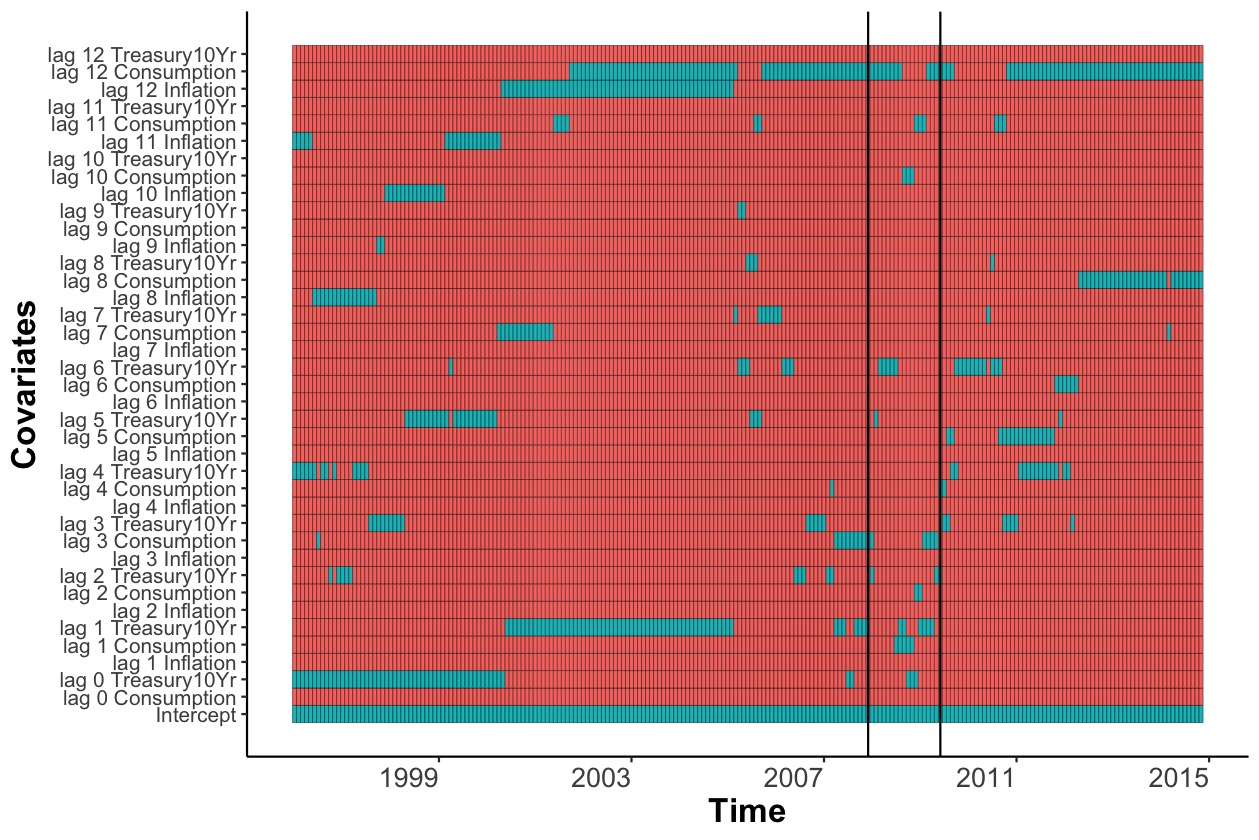}
%
\includegraphics[width=.75\textwidth]{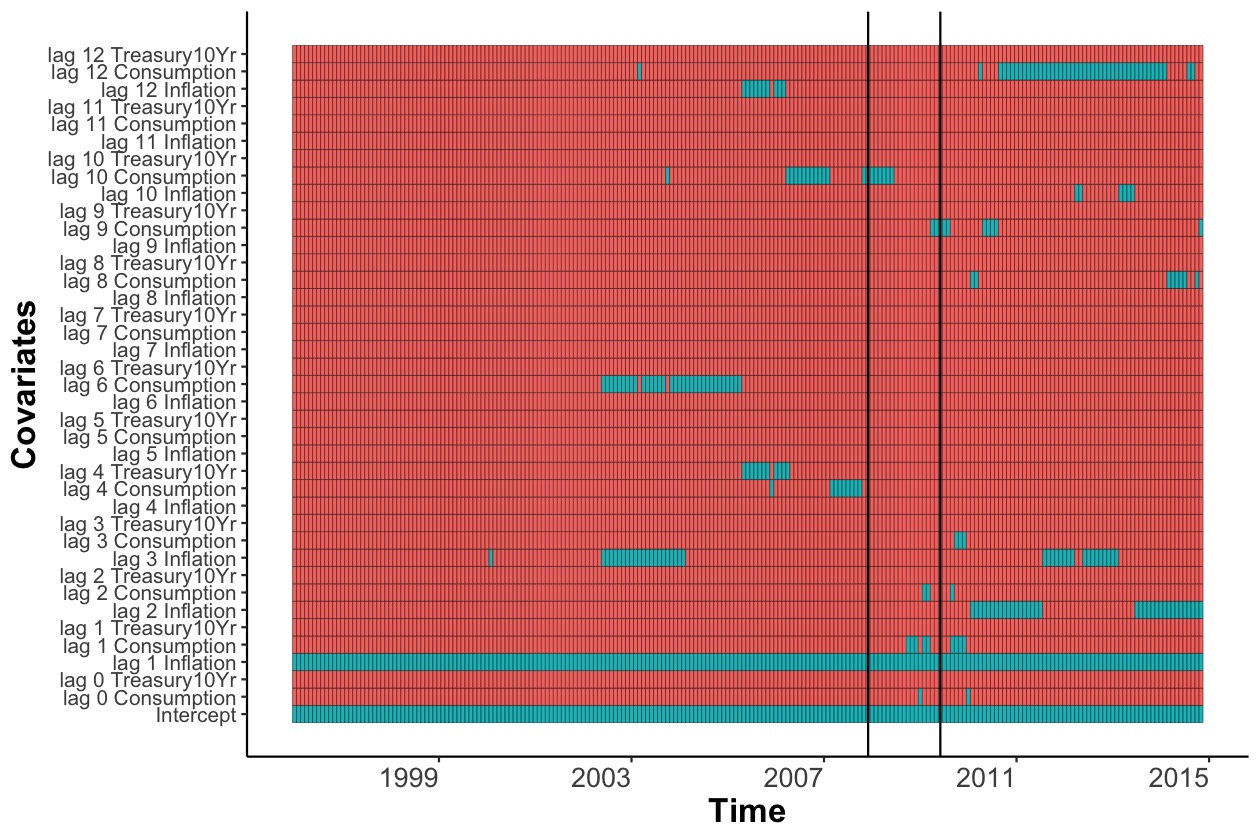}
\caption{Macroeconomic example:  Marginal $k = 24-$steps ahead forecasting comparisons. Dynamic variable inclusion ({\em green}) for DDNM series $j=1$, Inflation, in the posterior modal models:  
under marginal $24-$step AVS ({\em upper}) and BMA ({\em lower}). Vertical bars indicate the start and end of the great recession, during which AVS rapidly adapts to changing economic conditions.
 \label{fig-t_plus_k_density_avsbma_model} }
\end{figure}

Differences between  model weightings and selection under AVS and BMA  can be visualized in terms of variables included in the modal DDNMs and how these variable sets change over time. For BMA this is simply the model with maximum posterior probability at each time point, while for AVS it is the representative modal model at each time.  The  DDNM component models for the Inflation series are highlighted in Figure~\ref{fig-t_plus_k_density_avsbma_model}.
AVS focuses on higher lags of predictor variables, particularly lag$-12$ Inflation and lag$-12$ Consumption. Models featuring these higher lags produce more stable and accurate longer-term forecasts, although they are less accurate for $1-$step forecasting than models which include the lag$-1$ variable that are more favored under BMA.

\FloatBarrier

\subsection{Multi-step Path Forecasting \label{ksteppathforecasting}}

A major interest lies in improved {\em path forecasting}. In this applied context, the main focus is on how the macroeconomy is predicted to evolve over a coming period of months, and how the series are predicted to interact over that time.  Such goals are naturally addressed in the Bayesian framework by exploring full joint predictive distributions  over multiple months.  This is contrasted with the usual horizon-specific, or marginal forecasting analysis, of Section~\ref{kstepmargin}.  With a focus on the path over the next $k$ time points, 
we are therefore interested in the ($k\times m-$dimensional) {\em   path forecast density} $p_t(\by_{t+1},\ldots,\by_{t+k} | \cD_t)$, and refer  to understanding the underlying distribution as  path forecasting.  The suffix $t$ makes explicit that this is the joint forecast over the next $k$ time points made at time $t.$ 
In addition to potentially extracting distributional summaries,  one key use of models is simulation:  generating \lq\lq synthetic future paths'' of the economy that can be explored subjectively and used to interrogate predictions on arbitrary functions of the economic variables (e.g., defined downturns, etc).   In terms of model structure assessment, the explicit aim is to find models that balance short and longer-term forecasting, rather than focus on one or more specific horizon. 
 
Define the corresponding log path forecast density (LPFD) score at any time $t$  via
\begin{equation} \label{LPFDS}
s_t(\cM) = \sum_{h=0}^{t-k} \alpha^{t-k-h} \log ( p_h(\by_{h+1},\ldots,\by_{h+k} | \cM, \cD_h) )
\end{equation}
with discount $\alpha\in (0,1]$. As above, the example sets $\alpha = 0.98$ for both AVS and BMA.
As the loss function is based on a $k$-dimensional joint density, the natural setting for the scale parameter $\tau$ in Gibbs model probabilities is $\tau=1/k$; this puts the model score on the same scale as in standard Bayesian updating of model probabilities. As in Section \ref{kstepmargin}, the computation scales linearly in the number of series because the LPFD score can be factored into independent terms for each univariate series.


In any chosen DDNM, forecast evaluation of path scores is  via Monte Carlo.  This involves simple, direct/forward simulation of  state vectors and volatilities in the usual recursive form within each time point, and then sequentially over the next $k$ time points.   This generates samples from the predictive distributions of these latent parameter processes, each of which defines a full set of conditional multivariate normal distributions for the outcome path 
$\by_{h+1},\ldots,\by_{h+k}.$     Monte Carlo averages of these normals evaluated at the eventual outcome data provide Monte Carlo evaluations of path densities.    

\subsection{Path Forecasting Results \label{pathforecastingresults} }

\begin{figure}[h!] 
\centering
\includegraphics[width=12 cm]{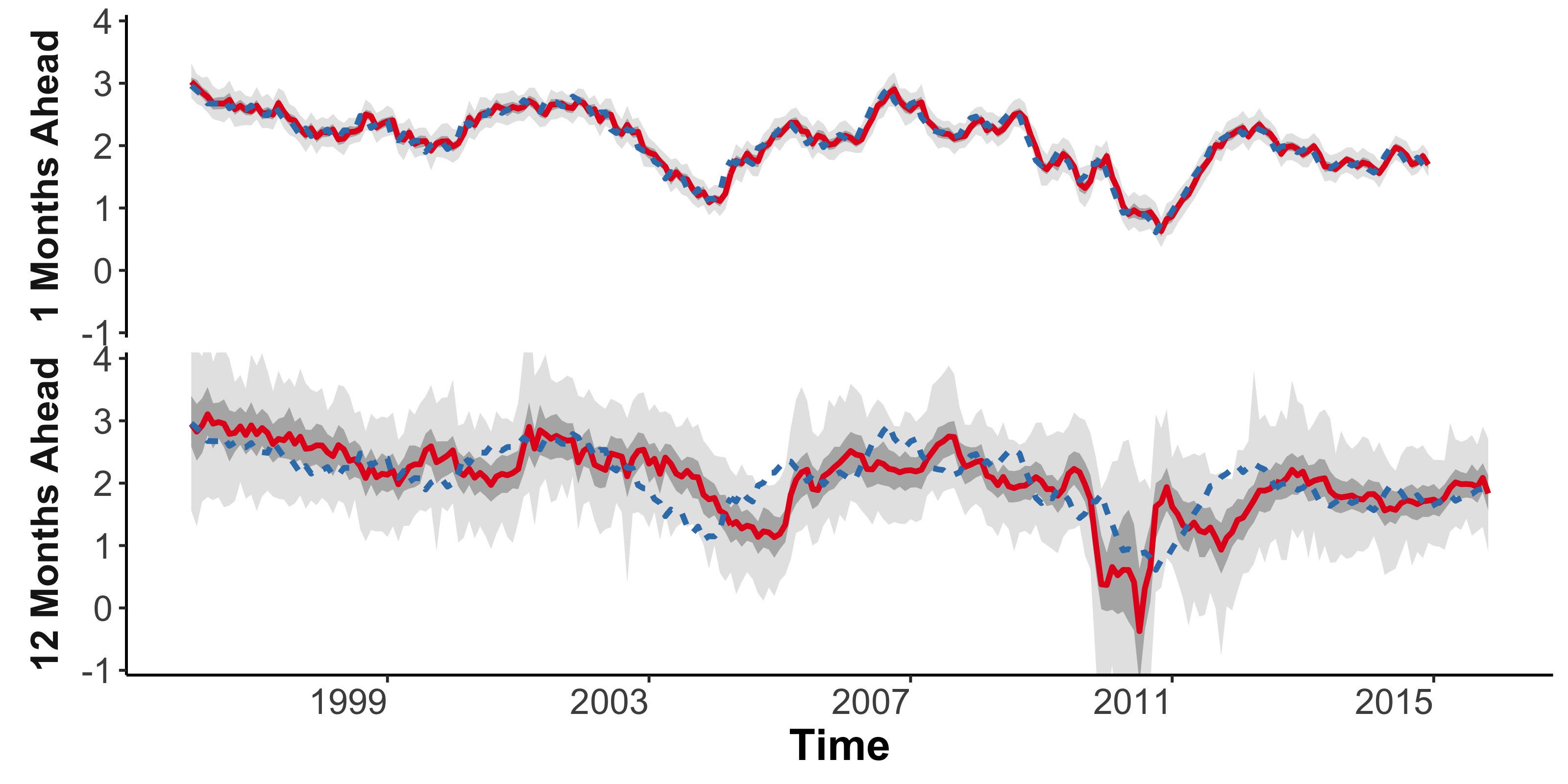}
\includegraphics[width=10 cm]{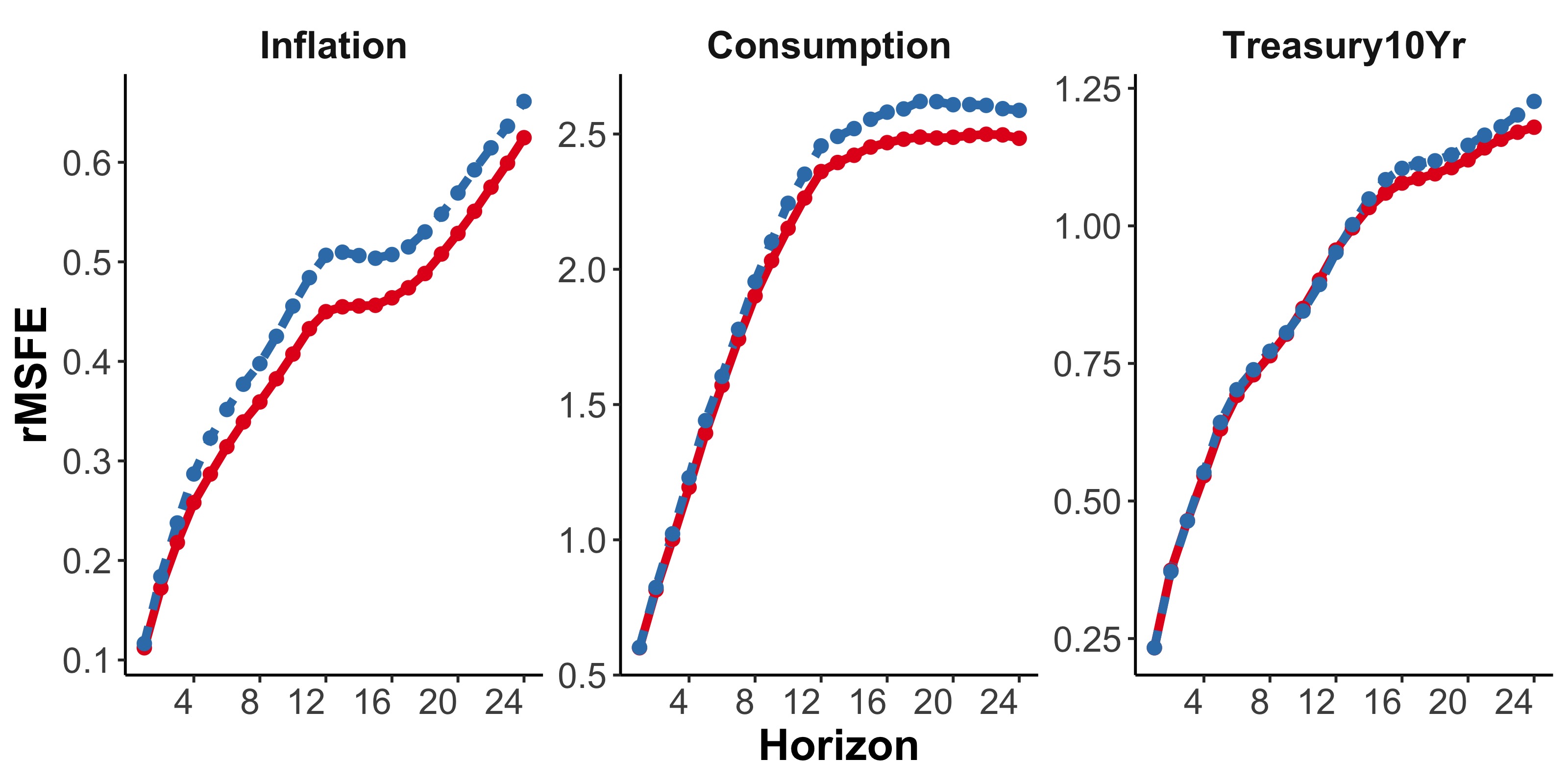}
\caption{Macroeconomic example: Path forecasting comparisons. The upper frame shows $1-$month and $12-$months ahead forecasts of Inflation using AVS:  data ({\em blue, dashed}), forecast means ({\em red, solid}), $50\%$ prediction intervals ({\em dark gray bands})  and $95\%$ prediction intervals ({\em light gray bands}).   The lower frames show
marginal rMSFE measures for Inflation, Consumption, and Tr10Yr over 1 to 24 month forecast horizons for AVS ({\em red, solid}) and BMA ({\em blue, dashed}). 
\label{fig-macro-Inflation-forecast-rMSFE}  }
\end{figure}

The expectation is that path forecast-guided AVS will  improve longer-term (up to horizon $k$)  forecast accuracy while still favoring  models with realistic short-term forecasts. This is borne out. Figure~\ref{fig-macro-Inflation-forecast-rMSFE} 
shows summary information for $1-$ and $12-$month ahead Inflation forecasts.  Related plots for Consumption and Tr10Yr are in Appendix 2 
of the Supplement. 
Improved forecasting accuracy under AVS can be seen in the marginal rMSFEs. 
The trend is for the short-term forecast accuracy to be very similar to BMA, with AVS offering increasing improvements over BMA at longer forecast horizons.    

\begin{figure}[hpt!] 
\centering
\includegraphics[width=.75\textwidth]{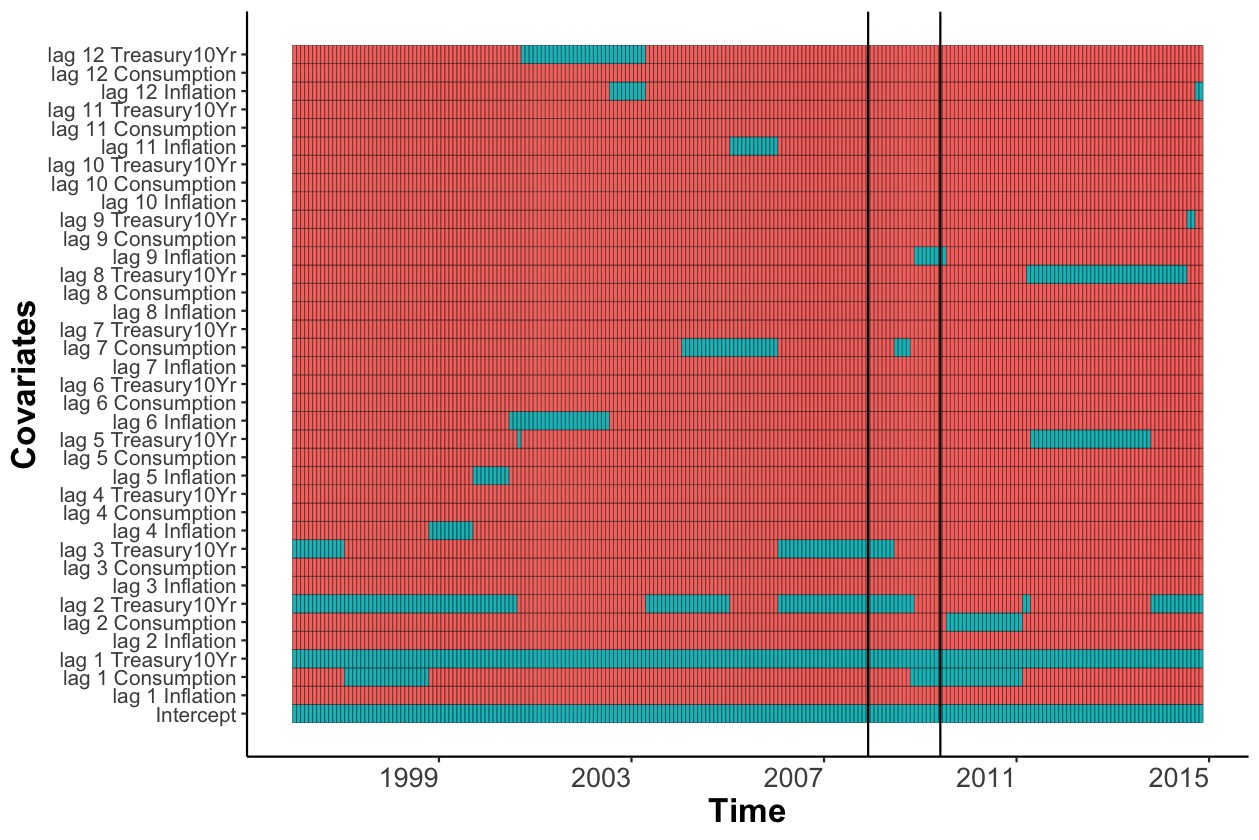}
%
\includegraphics[width=.75\textwidth]{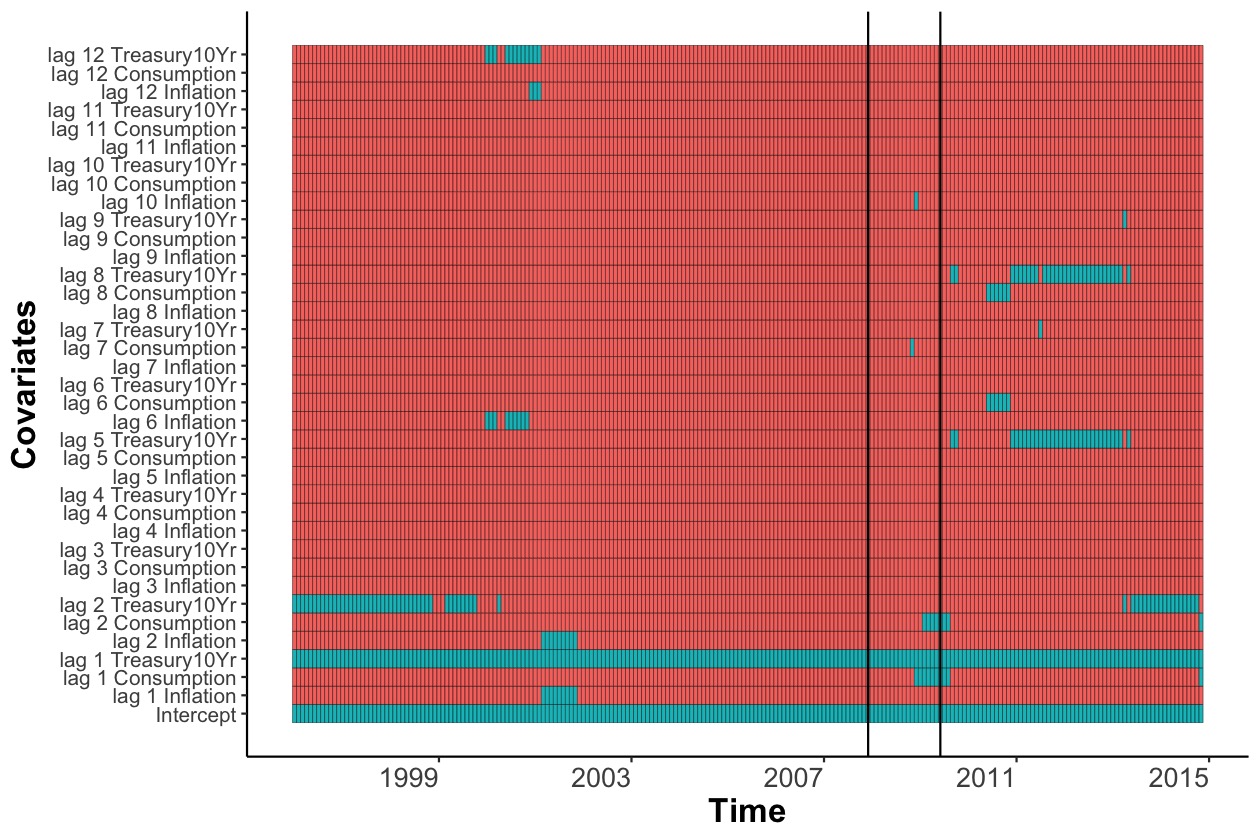}
\caption{Macroeconomic example: Path forecasting comparisons. Dynamic variable inclusion ({\em green}) for DDNM series $j=3$, Tr10Yr, in the posterior modal models:  
under  $24-$step path forecasting AVS ({\em upper}) and BMA ({\em lower}). Vertical bars indicate the start and end of the great recession, during which AVS focuses on shorter term predictors to adapt to changing economic conditions.
 \label{fig-model-tr10yr-avsbma} }
\end{figure}

Trajectories over time of indicators of variables included in modal models provide insights into differences between AVS and BMA in this path forecasting context; this is  illustrated in Figure~\ref{fig-model-tr10yr-avsbma}  in DDNM model components for predicting series $j=3,$ Tr10Yr.
It is typical and to be expected that the lag$-1$ value of a given series is a dominant predictor of that series, especially with data at monthly levels. 
This holds true in the models selected by both BMA and AVS for all 3 series, exemplified in this figure for Tr10Yr. 
However, with the LPFD score using $k=24$, AVS does better in capturing longer-term dynamics; the figure highlights the involvement of higher lags of all series in the AVS analysis relative to that using standard BMA. 


\subsection{Higher Dimensional Results \label{higherdimresults} }
As described in Sections \ref{kstepmargin} and \ref{ksteppathforecasting}, the computational load of AVS analysis in DDNMs  scales {\em linearly} in the number of series, enabled by evaluating model scores independently across series in the DDNM hierarchy.  A further study of a higher-dimensional series confirms the ability to scale computations while also reinforcing the role of the AVS strategy in goal-focused forecasting.


We expand the previous analysis to $7-$dimensions by adding time series of year-over-year Wage Growth, M2 Money Stock, Moody's BAA Corporate Bond, and Gold prices. These variables are important indicators of the labor market, monetary policy, and corporate activity. We consider as potential predictors 1, 3, 6, and 12 month lags of each series. The model score is the marginal $k=24$ month ahead log predictive density, as defined in Section \ref{kstepmargin}. As in the previous example, we set $\alpha=0.98$ for both AVS and BMA, and $\tau=1$ to put the model score on the same scale as standard Bayesian model probabilities.

\begin{figure}[htbp!]
	\centering
	\includegraphics[width=0.8\textwidth]{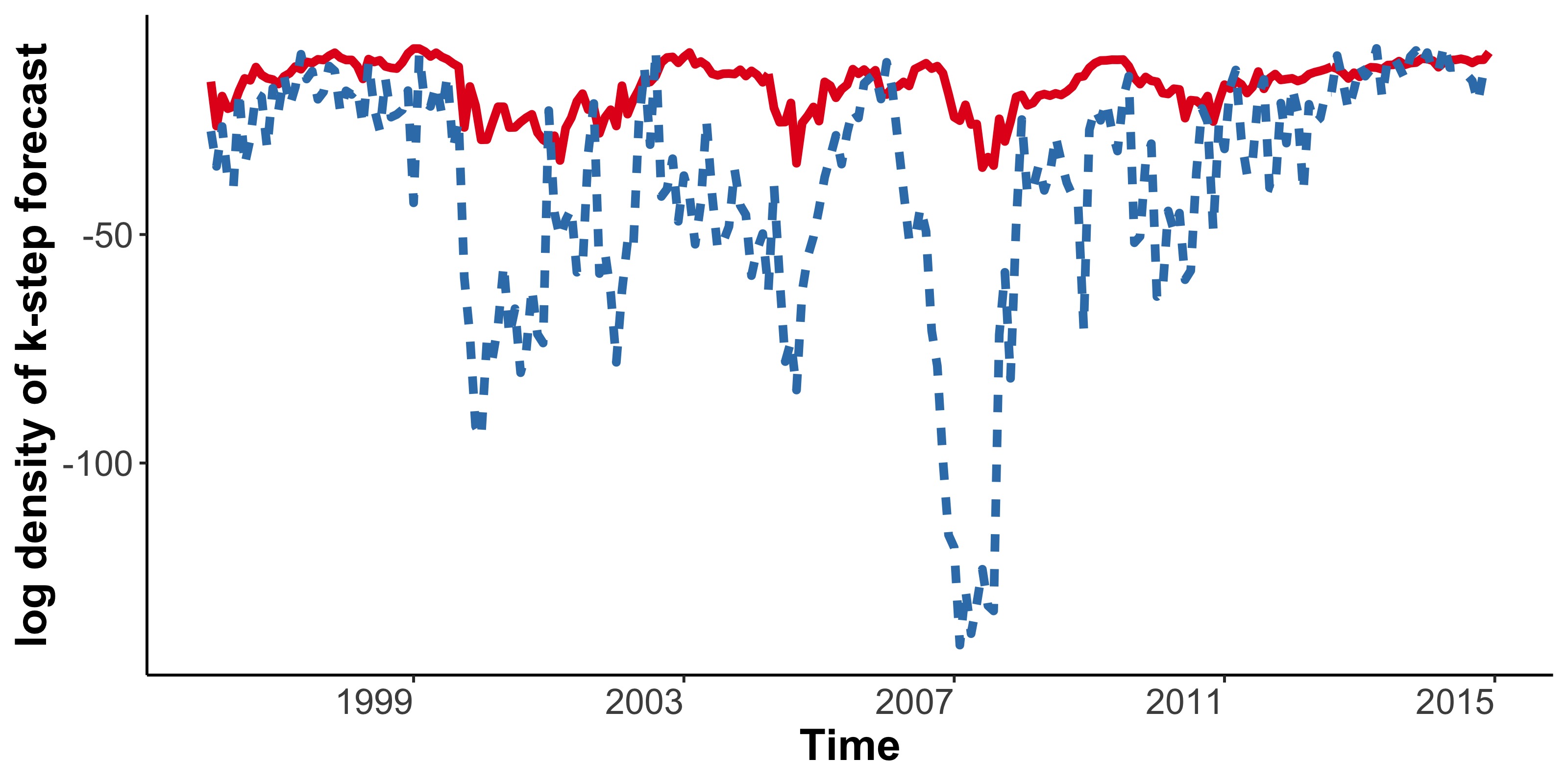}

	\caption{$7-$Dimensional macroeconomic example: Log forecast densities $\log (p(\by_{t+24}|\cD_t ))$ over time $t$ for AVS ({\em red, solid}) and BMA ({\em blue, dashed}).}
		
	\label{fig-t_plus_k_higherdim_forecastperformance}
\end{figure}

\begin{figure}[htbp!]
	\centering
	\includegraphics[width=0.8\textwidth]{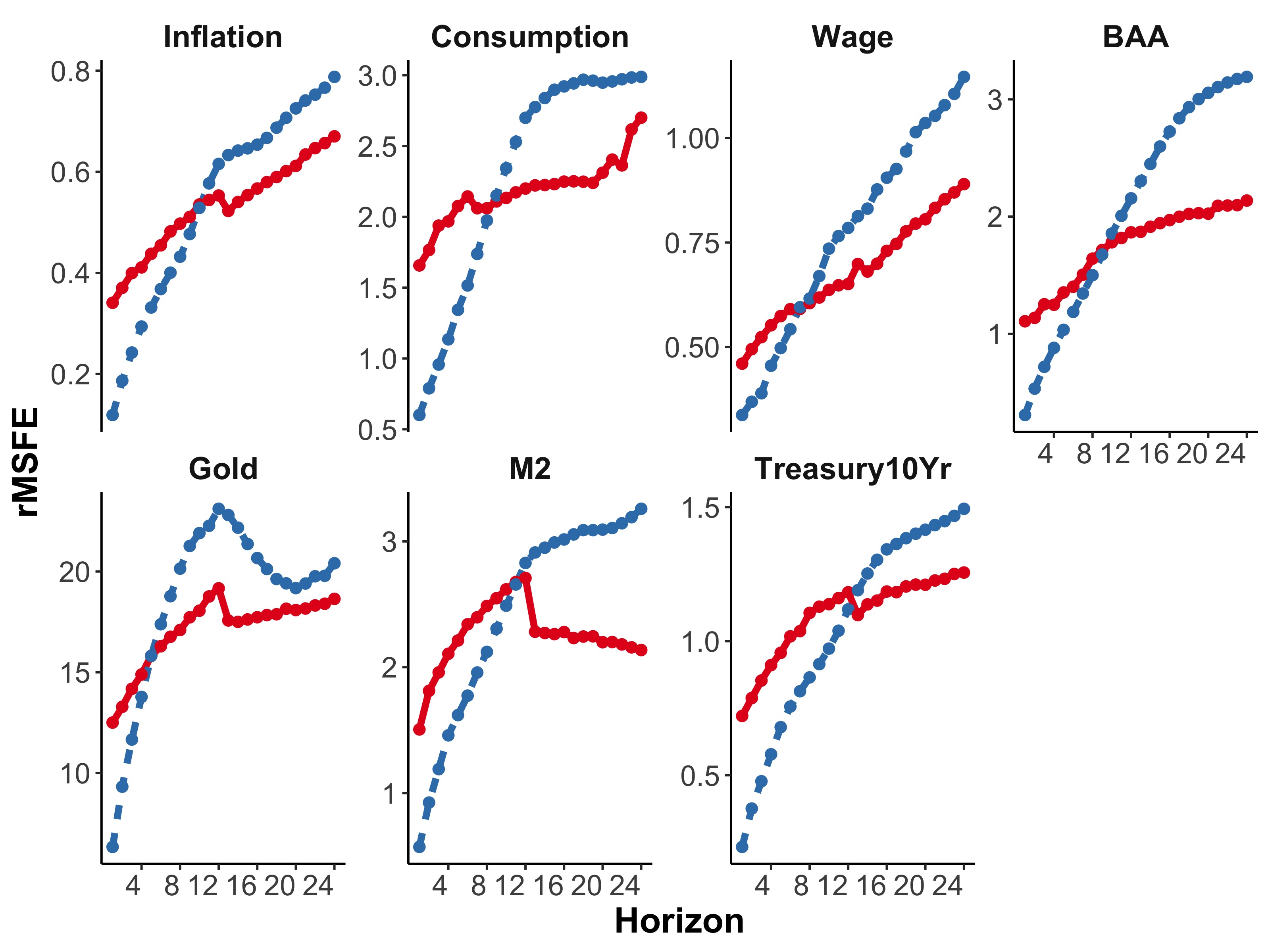}
	
	\caption{$7-$Dimensional macroeconomic example: Marginal rMSFE measures for all series over 1 to 24 month forecast horizons for AVS ({\em red, solid}) and BMA ({\em blue, dashed}).}
	
	\label{fig-t_plus_k_higherdim_forecastmse}
\end{figure}

Figures \ref{fig-t_plus_k_higherdim_forecastperformance} and \ref{fig-t_plus_k_higherdim_forecastmse} compare the performance of AVS and BMA. The higher-dimensional DDNM provides a greater diversity of economic signals to choose from as predictors, leading to larger differences between AVS and BMA. As in the $3-$dimensional example, AVS consistently dominates BMA with respect to the $24-$month ahead log forecast density, especially during periods of economic upheaval. The marginal rMSFE shows that BMA performs better at short-term forecasting, while AVS produces superior results over longer horizons.

\section{Additional Comments} \label{conclusions}

\subsection{Conclusions} 
AVS builds on concepts   in the recent Bayesian literature to define goal-oriented model structure uncertainty analysis that avoids the shortcomings of standard approaches. We do this in sequential, dynamic time series contexts by adapting  goal-focused Gibbs model probabilities coupled with efficient shotgun stochastic search over spaces of model structures, overlaid on standard Bayesian analysis in DDNMs.   The resulting methodology maximally exploits analytic Bayesian computations within DDNMs, and is open to partial parallelization of both analytic and direct simulation-based computations  for forecasting for increasingly high-dimensional series. 
 
The examples with simulated and real economic data show the ability of goal-focused AVS to  achieve superior results to standard Bayesian model structure learning.   AVS improves longer-term forecast accuracy by identifying, weighting and averaging over models whose structure is different to that identified by standard Bayesian analysis;  time series models more heavily weighted for longer-term forecasting naturally involve longer-lagged predictors. The important context of path forecasting emphasizes the benefits of AVS while also highlighting the relevance of standard Bayesian model probability analysis in connection with shorter-term forecasting. 

\subsection{Potential Extensions and Related Research}
  
The paper uses DDNMs as context and examples.     The related class of 
Simultaneous Graphical Dynamic Linear Models (SGDLMs)  relax  restrictions on the selection of parental predictors~\citep{GruberWest2016BA,GruberWest2017ECOSTA}, with a trade-off in terms of additional computational needs. AVS is  directly extensible to SGDLMs (and, to other multivariate dynamic model frameworks, at least in principle) and further development in that direction can be anticipated.

The presentation and examples in the paper focus on the use of model scores based on a specific, defined forecasting or decision goal.    In other settings, there may be several-- possibly competing-- goals.  For example, in multi-step forecasting we may consider marginal forecasts at each horizon $h=1,2,\ldots k$ to be of explicit interest. Fitting separate models and AVS analyses for each horizon would be is consistent with \lq\lq models for goals'' as in Bayesian predictive synthesis approaches to model combination~\citep{McAlinnWest2017bpsJOE,McAlinnEtAl2017,McAlinnEtAldiscussionBA2018}, and with the over-arching motivation for AVS. This  obviously raises questions of  computational demands  as well as of how to balance and potentially combine AVS analyses across goals.    An alternative view is to use some form of aggregate score that balances interest across the several goals, consistent with practice in multi-objective Bayesian decision analysis.

\newpage

\subsection*{Appendix A: Univariate DLMs} \label{appendix-univariate-dlm}

As noted in in Section 2 
the set of univariate DLMs adopted for the DDNM components are standard models in which the state vectors $\btheta_{j,t}$ and volatilities (precisions) $\lambda_{j,t}$ evolve jointly according to random walks, and independently across series $j.$  See details and standard  material for DLMs in chapt. 4 in each of~\cite{WestHarrison1997} and~\cite{PradoWest2010}.  For additional details here, we note the following summaries of technical components of prior, posterior and forecast distributions involved in the basic Bayesian computations.

\textbf{Posteriors at time $t-1$:} Independently for each series $j$,  we have normal-gamma posteriors for $(\btheta_{j,t},\lambda_{j,t})$, viz
\begin{equation*}
\begin{split}
\btheta_{j,t-1} | \lambda_{j,t}, \cD_{t-1} & \sim N(\bm_{j,t-1}, \bC_{j,t-1}/({s_{j,t-1} \lambda_{j,t-1}})), \\
\lambda_{j,t-1} | \cD_{t-1} & \sim Ga({n_{j, t-1}}/{2}, {n_{j, t-1} s_{j, t-1}}/{2}).
\end{split}
\end{equation*}

\textbf{Priors at time $t$:} Posteriors at time $t-1$ evolve to priors at time $t$ via evolution equations 
\begin{equation*} \label{state_evolution}
\begin{split}
\btheta_{j,t} &= \btheta_{j, t-1} + \bomega_{j,t} \quad \text{where} \quad \bomega_{j,t} \sim N(0, \bW_{j,t}/({s_{j,t-1} \lambda_{j,t}})), \\
\lambda_{j,t} &= \lambda_{j,t-1} {\eta_{j,t}}/{\beta_j} \quad \text{where} \quad \eta_{j,t} \sim Be({\beta_j n_{j,t-1}}/{2}, {(1-\beta_j)n_{j,t-1}}/{2}),
\end{split}
\end{equation*}
and where $\bW_{j,t}=\bC_{j,t-1}(1-\delta_j)/\delta_j$  is defined by a single discount factor $\delta_j \in (0, 1]$, and the independent beta random variables $\eta_{j,t}$ are defined by a discount factor $\beta_j \in (0, 1]$.   This results in priors at time $t$ given by 
\begin{equation*}
\begin{split}
\btheta_{j,t} | \lambda_{j,t}, \cD_{t-1} & \sim N(\ba_{j,t}, \bR_{j,t}/(s_{j,t} \lambda_{j,t})), \\
\lambda_{j,t} | \cD_{j, t-1} & \sim Ga({n_{j, t-1}}/{2}, {n_{j, t-1} s_{j, t-1}}/{2})
\end{split}
\end{equation*}
with $\ba_{j,t} = \bm_{j,t-1}$, $\bR_{j,t} = \bC_{j,t-1} / \delta_j$ and $r_{j,t} = \beta_j n_{j, t-1}$.

\textbf{Forecasting $1-$step ahead:} The  predictive distribution for series $j$ is a univariate $t-$distribution
\begin{equation}
y_{j,t} | \by_{pa(j),t}, \cD_{t-1} \sim T_{r_{j,t}} \left(\bF'_{j,t} \ba_{j,t},  s_{j, t-1} + \bF'_{j,t} \bR_{j,t} \bF_{j,t} \right)
\end{equation}

\newpage
\subsection*{Appendix B: AVS Forecasts} \label{appendix-avs-forecasts}

\begin{figure}[htbp!] 
\centering
\includegraphics[width=12 cm]{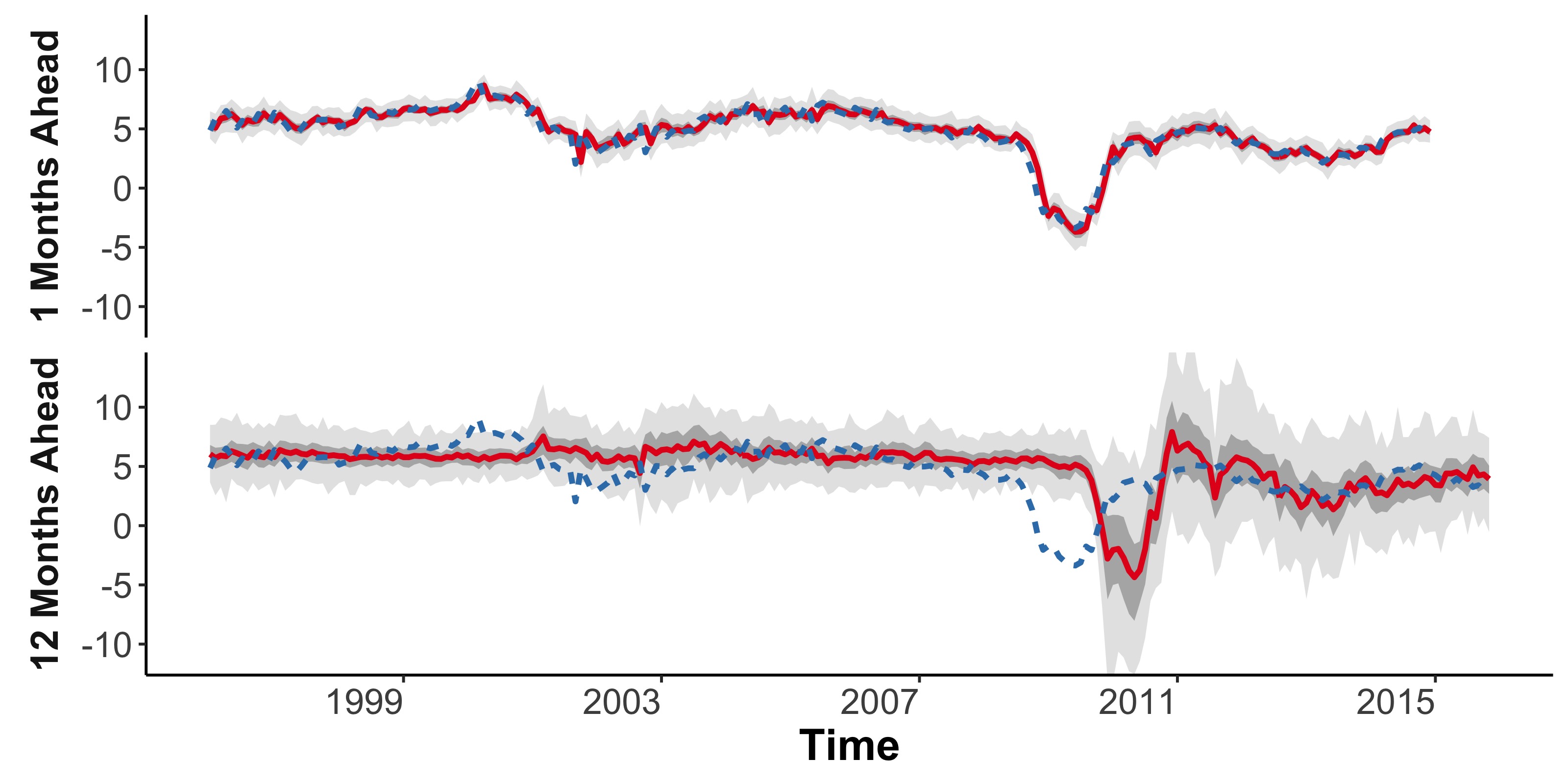}
%
\includegraphics[width=12 cm]{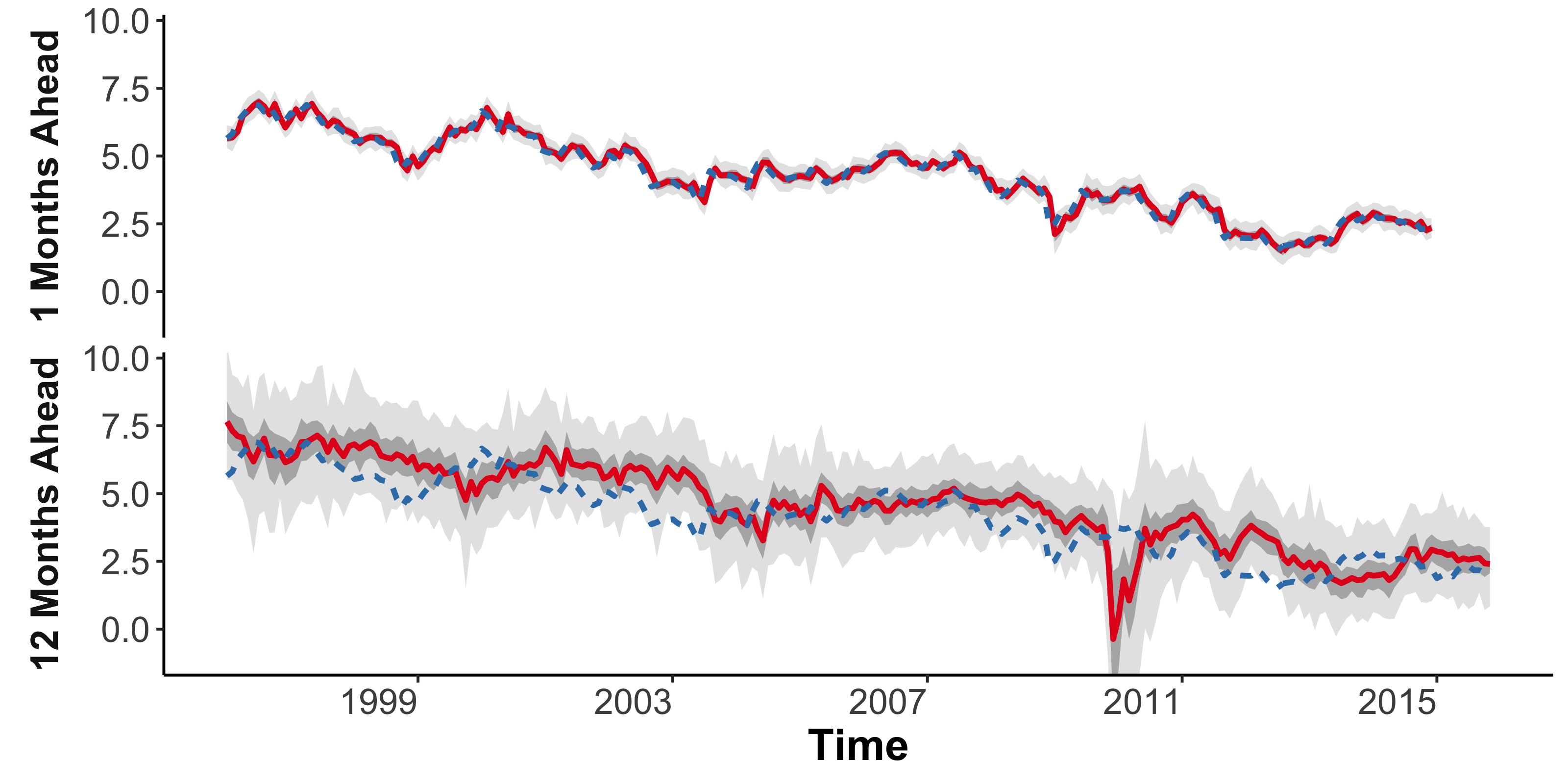}
\caption{Macroeconomic example: Path forecasting comparisons. $1-$month and $12-$months ahead forecasts for Consumption ({\em upper}) and Tr10Yr  ({\em lower}) using AVS, showing  data ({\em blue}), forecast means ({\em red}), $50\%$ prediction intervals ({\em dark gray bands})  and $95\%$ prediction intervals ({\em light gray bands}).    
\label{fig-macro-Consumption+Tr10Yr-forecast}  }
\end{figure}

Referring to Section 5.4 and 
 Figure 7, 
 the corresponding plots of forecast summaries for the 
Consumption and Tr10Yr series under the AVS analysis are given in  Figure ~\ref{fig-macro-Consumption+Tr10Yr-forecast}. 

\newpage

\subsubsection*{\bf Acknowledgements}

The research reported here was developed while Isaac Lavine and Michael Lindon were PhD students in Statistical Science at Duke University. 
Aspects of the research developed in the working group on  {\em Bayesian Analysis \& Decisions} of the 2016-2017 program on {\em Optimization} of the Statistical and Applied Mathematical Sciences Institute (SAMSI).   Isaac Lavine and Michael Lindon  were the co-recipients of the 
2016-17  BEST Award for Student Research at Duke University, and acknowledge the BEST Foundation for partial financial support in early stages of the research reported here.

\bibliography{AVSarXiv2020}
\bibliographystyle{chicago}

\end{document}